\documentclass{IEEEtran}
\IEEEoverridecommandlockouts
\usepackage{amsmath,amssymb,amsfonts}
\usepackage{algorithm}
\usepackage{algorithmic}
\usepackage{bigstrut}
\usepackage{graphicx}
\graphicspath{{./Pictures/}} 

\usepackage{cite}
\usepackage{textcomp}
\ifCLASSOPTIONcompsoc
\usepackage[caption=false,font=normalsize,labelfont=sf,textfont=sf]{subfig}
\else
\usepackage[caption=false,font=footnotesize]{subfig}
\fi
\usepackage{xcolor}

\usepackage{array}
\newcolumntype{L}[1]{>{\raggedright\let\newline\\\arraybackslash\hspace{0pt}}m{#1}}
\newcolumntype{C}[1]{>{\centering\let\newline\\\arraybackslash\hspace{0pt}}m{#1}}
\newcolumntype{R}[1]{>{\raggedleft\let\newline\\\arraybackslash\hspace{0pt}}m{#1}}

\newcommand{\tabincell}[2]{\begin{tabular}{@{}#1@{}}#2\end{tabular}} 
\def\BibTeX{{\rm B\kern-.05em{\sc i\kern-.025em b}\kern-.08em
    T\kern-.1667em\lower.7ex\hbox{E}\kern-.125emX}}
\begin{document}

\title{Adaptive Quantization for Key Generation in Low-Power Wide-Area Networks}
\author{Chen Chen, \textit{Member, IEEE},  Junqing~Zhang, \textit{Member, IEEE}, and Yingying Chen, \textit{Fellow, IEEE}

\thanks{Manuscript received xxx; revised xxx; accepted xxx. Date of publication xxx; date of current version xxx. The work was in part supported by UK EPSRC under grant ID EP/V027697/1 and the National Key Research and Development Program of China under grant ID 2020YFE0200600. The review of this paper was coordinated by xxx. 	\textit{(Corresponding author: Junqing Zhang.)}}
\thanks{Chen Chen and Junqing Zhang are with the Department of Electrical Engineering and Electronics, the University of Liverpool, Liverpool, L69 3GJ, UK. (email: c.chen77@liverpool.ac.uk; Junqing.Zhang@liverpool.ac.uk)
}
\thanks{Yingying Chen is with the Department of Electrical and Computer
Engineering and the Wireless Information Network Laboratory,
Rutgers University, New Brunswick, NJ 08854 USA (e-mail:
yingche@scarletmail.rutgers.edu).
}
\thanks{Color versions of one or more of the figures in this paper are available online at http://ieeexplore.ieee.org.}
\thanks{Digital Object Identifier xxx}	
}

\maketitle

\begin{abstract}
 Physical layer key generation based on reciprocal and random wireless channels has been an attractive solution for securing resource-constrained low-power wide-area networks (LPWANs). When quantizing channel measurements, namely received signal strength indicator (RSSI), into key bits, the existing works mainly adopt fixed quantization levels and guard band parameters, which fail to fully extract keys from RSSI measurements. In this paper, we propose a novel adaptive quantization scheme for key generation in LPWANs, taking LoRa as a case study. The proposed adaptive quantization scheme can dynamically adjust the quantization parameters according to the randomness of RSSI measurements estimated by Lempel-Ziv complexity (LZ76), while ensuring a predefined key disagreement ratio (KDR). Specifically, our scheme uses pre-trained linear regression models to determine the appropriate quantization level and guard band parameter for each segment of RSSI measurements. Moreover, we propose a guard band parameter calibration scheme during information reconciliation during real-time key generation operation. Experimental evaluations using LoRa devices show that the proposed adaptive quantization scheme outperforms the benchmark differential quantization and fixed quantization with up to 2.35$\times$ and 1.51$\times$ key generation rate (KGR) gains, respectively. 
\end{abstract}


\section{Introduction}
The past decades have witnessed the exponential growth of Internet of Things (IoT) devices \cite{Survey1, Survey2}. IoT has improved the quality of our lives and contributed to the growth of the economy with its wide applications in smart homes, smart transportation, healthcare, industrial automation, etc. Recently, low power wide area networks (LPWANs) have gained popularity thanks to their ability to provide long-distance communications for devices in a wide area. Typical LPWAN techniques include LoRa, Sigfox, and narrowband IoT (NB-IoT). The widespread application of LPWANs has brought massive convenience and benefits, but it also increases security vulnerability due to the broadcast nature of wireless communications.

To secure IoT communications, data encryption is an integral step. For example, advanced encryption standard (AES) is adopted in LoRaWAN~\cite{lorawan11}, which is well defined. However, the provision of the symmetric key is not specified in the LoRaWAN specification but its implementation is left to users~\cite{lorawan11}. Computer networks use public key cryptography (PKC) to distribute keys for symmetric encryption, but PKC is rather complicated and computationally expensive for low-cost IoT applications~\cite{Access1}. 
In addition, PKC relies on complicated mathematical problems such as discrete logarithms; these algorithms cannot be scaled and will be threatened by the emerging quantum computer~\cite{cheng2017securing}.
In practice, many IoT devices are pre-programmed with a static key that is never refreshed, which poses security risks as they might be stolen or cracked.

In this context, physical layer key generation based on wireless channels becomes a promising technique, which exploits the  reciprocal and time-varying channel characteristics to generate secure keys \cite{Access1, WC1}.  More importantly, it is a lightweight technique with much lower computational complexity and energy consumption compared to PKC-based key distribution mechanisms. 
It has been proved information-theoretically secure by two pioneering and seminal papers in 1993~\cite{TIT2, TIT3}.
The theoretical foundation inspired many practical explorations in commercial communication networks; the majority of the existing key generation work focuses on short-range communication techniques, such as WiFi~\cite{Mobicom,wei2011adaptive,TMC1,Wifi,zhang2016experimental}, ZigBee~\cite{Zigbee} and Bluetooth~\cite{Bluetooth}. 

Compared to short-range communications-based key generation efforts, there is limited work investigating key generation in LPWANs~\cite{IOTJ2,IOTJ1}. Ruotsalainen \textit{et al.} \cite{IOTJ2} investigated key generation performance for both LoRa signaling and LoRaWAN. They considered static mode where both legitimate nodes remain stationary and enhanced the channel randomness using an electrically steerable parasitic array of radiators (ESPARs) type antenna. Single-bit quantization was exploited to convert received signal strength indicator (RSSI) values into key bits. Xu \textit{et al.}~\cite{IOTJ1} conducted extensive RSSI measurements for different environments (outdoor and indoor), modes (static and mobile), distances, and data rates. A key generation protocol was proposed, which employed multi-level quantization and compression sensing-based information reconciliation. 

The aforementioned  works selected fixed quantization parameters according to the collected RSSI measurements, which may not lead to optimal key generation performance when the channel diversity varies significantly under different moving patterns and environmental changes. 
Hence, fixed quantization parameters can no longer adapt to different channel diversities.
Zhang \textit{et al.} \cite{TVT1} showed that the RSSI measurements may experience large variations, from $-$123 dBm to $-$49 dBm, in a test conducted in an urban environment. Differential value-based quantization was adopted, which can capture the RSSI variations in an adaptive manner by comparing the adjacent RSSI values. However, it generates no more than 1 key bit per channel measurement, which significantly limits the key generation rate (KGR). 

There have been several works tackling  multi-level quantization for key generation \cite{ ye2006extracting, ye2010information, patwari4967595, TMC2, li2022fast, jiao2021machine}. In \cite{ye2006extracting, ye2010information}, Ye \textit{et al.} showed that a higher quantization level leads to a higher secrecy rate at the cost of a higher bit error rate based on theoretical analysis and simulations. In \cite{patwari4967595}, Patwari \textit{et al.} proposed a  multi-level  quantization method, where the channel measurements are divided into equally likely quantization levels by calculating the cumulative distribution function (CDF). In \cite{TMC2}, Premnath \textit{et al.} partitioned the channel measurements into small blocks and performed multi-level  quantization in each block. During quantization, the range of channel measurements was divided into equal-sized intervals and each interval was assigned one or multiple key bits. The experimental results revealed that higher quantization levels are more suitable for mobile scenarios than stationary ones. In \cite{li2022fast}, Li \textit{et al.} investigated physical layer key generation in slowly varying environments using channel obfuscation. A multi-level  quantization method similar to that in \cite{TMC2} was adopted to extract key bits from the channel measurements. Nevertheless, these works only presented the effect of quantization level on KGR and key disagreement ratio (KDR) based on the channel measurements of both legitimate nodes. Although \cite{patwari4967595, TMC2} have reported that different channel variations require different optimal quantization levels,  how to determine the optimal quantization level has not been explored. As such, the aforementioned  multi-level quantization methods cannot be applied to  varying channel diversities.  Recently, Jiao \textit{et al.} studied reconfigurable intelligent surface (RIS)-assisted key generation, and proposed a machine learning-based adaptive quantization scheme to predict the optimal quantization level as per the channel state information observed.  However, the randomness of the generated keys was not taken into account and only simulation results were provided. Its performance under practical scenarios remains unknown.

In order to fully exploit the randomness from diverse channel variations, this paper designed an adaptive multi-level quantization approach that can dynamically select the optimal quantization parameters. Specifically, we trained linear regression models to learn the mapping between channel randomness and optimal quantization parameters. The developed linear regression models can be used to adaptively select quantization parameters as per channel diversity.  The proposed approach was evaluated by comprehensive experiments using a LoRa testbed.
The main contributions of this paper are summarized as follows.
\begin{itemize}
\item We design a novel adaptive multi-level quantization scheme that can automatically select appropriate quantization levels and  guard band parameters according to channel variations. The quantizer is trained offline using pre-collected RSSI measurements.
\item We deploy the pre-trained quantization scheme in an adaptive quantization-based key generation (AQ-KG) protocol that can operate in real-time.
Since the selected quantization parameters may still fail when the channel variations deviate significantly from the training stage, a calibration method is accordingly designed to further tune the guard band parameters.
\item We conduct extensive real-world  channel measurements using a LoRa testbed to evaluate  the proposed AQ-KG protocol in urban, suburban, and indoor environments. Experimental results reveal that the proposed adaptive quantization scheme outperforms the state-of-the-art approaches, namely differential quantization and fixed quantization, with up to 2.35$\times$ and 1.51$\times$ KGR gains, respectively. 
The randomness of the generated keys is validated by the National Institute of Standards and Technology (NIST) test. The designed adaptive quantization scheme has demonstrated good generalization capability to {different environments and devices}.
\end{itemize}

The rest of this paper is organized as follows. In Section \ref{Preliminary}, we introduce the key generation primer. Section \ref{Problem Statement} presents the problem to be solved and the system overview. Then, we detail the design of the proposed adaptive quantization scheme in Section \ref{Adaptive Quantization} and the AQ-KG protocol in Section~\ref{sec:keygen}. Our experimental results and analysis are given in Section \ref{PERFORMANCE EVALUATION}. The security of the proposed AQ-KG protocol is analyzed in Section \ref{sec:security_analysis}. Finally, Section \ref{Conclusions} concludes this paper.

\section{Key Generation Primer}
\label{Preliminary}
\subsection{Protocol}
\begin{figure}[!t]
\centerline{\includegraphics[width=3.4in]{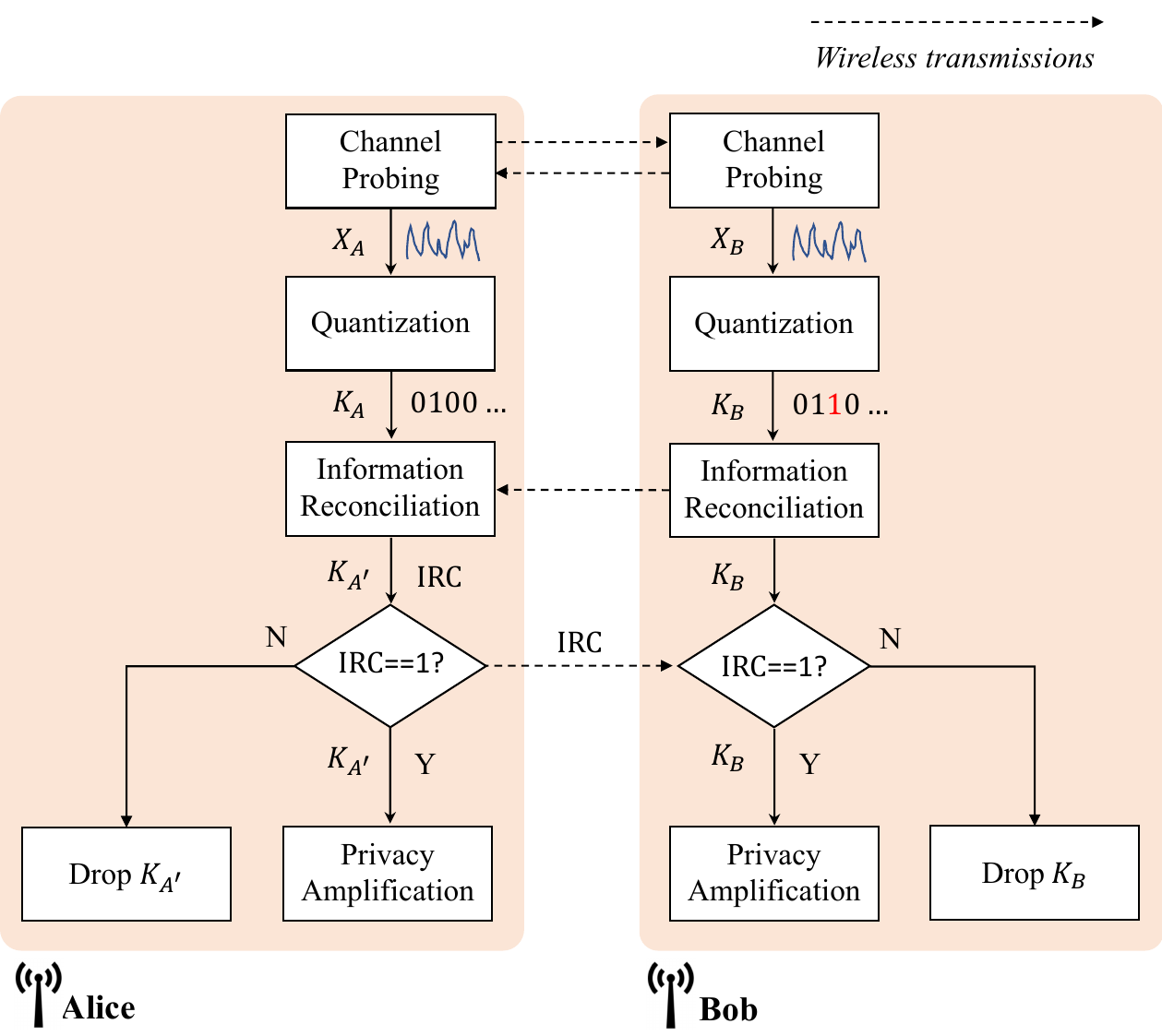}}
\caption{Workflow of a common key generation protocol.}
\label{fig:keygen_steps}
\end{figure}

Physical layer key generation exploits the noisy but highly correlated common channel between two terminals, Alice and Bob. As shown in Fig.~\ref{fig:keygen_steps}, a complete key generation protocol usually consists of four steps: channel probing, quantization, information reconciliation, and privacy amplification \cite{Access1}. 

\subsubsection{Channel Probing}
\label{sec:Channel_Probing1}
Channel probing performs bidirectional channel measurements between Alice and Bob.
After receiving a signal from Bob, Alice records the RSSI value $x_{A}$. Similarly, after receiving a signal from Alice, Bob records the RSSI value  $x_{B}$. These bidirectional channel measurements usually occur within the coherence time when the channel is considered to be reciprocal. 
After performing $N$ rounds of the above channel measurements, Alice and Bob obtain $N$ RSSI values, $X_{A} = \{x_{A}[1], x_{A}[2],\cdots,x_{A}[N]\}$ and $X_{B} = \{x_{B}[1], x_{B}[2],\cdots,x_{B}[N]\}$, respectively.

\subsubsection{Quantization}
Quantization converts the collected channel measurements into binary key bits. Multi-level quantization can extract multiple  bits from an RSSI measurement. For a sequence of RSSIs $X$, when the quantization level $m$, where $m\in \{2^{p}|p=1,2,3,\dots$\}, is adopted, the RSSIs are divided into the following $m$ intervals:
\begin{align}
Q_{1}=\big[q_{0}, q^{-}_{1}\big], Q_{2}=\big(q^{+}_{1}, q^{-}_{2}\big],
\dots, Q_{m}=\big(q^{+}_{m-1}, q_{m}\big]
\end{align}
where $q_{0}$ and $q_{m}$ are the minimum and maximum values of $X$, respectively, and
\begin{align}
q_{n}^{+} &=q_{n}+\alpha\times s_{n}\nonumber\\
q_{n}^{-} &=q_{n}-\alpha\times s_{n},	\nonumber
\end{align}
{where $q_n$ is the threshold for the $n$-th interval, $\alpha$ is the guard band parameter and $s_{n}=\mathbb{S}\left\{x\in X,  q_{n-1}< x\le q_{n+1}\right\}$, where $\mathbb{S}\{\cdot\}$ represents the standard deviation of a set.}
The calculation of $q_{n}$ obeys the following rules: given $q_{0}$ and $q_{m}$, $q_{\frac{m}{2}} = \mathbb{E}\{x\in X,  q_{0}\le x\le q_{m}\}$, where $\mathbb{E}\{\cdot\}$ denotes the mean of a set; 
\begin{itemize}
	\item When $m\ge4$, $q_{\frac{m}{4}} = \mathbb{E}\{x\in X,  q_{0}\le x\le q_{\frac{m}{2}}\}$ and $q_{\frac{3m}{4}} = \mathbb{E}\{x\in X,  q_{\frac{m}{2}}< x\le q_{m}\}$; 
	\item When $m\ge8$, $q_{\frac{m}{8}} = \mathbb{E}\{x\in X,  q_{0}\le x\le q_{\frac{m}{4}}\}$, $q_{\frac{3m}{8}} = \mathbb{E}\{x\in X,  q_{\frac{m}{4}}< x\le q_{\frac{m}{2}}\}$, $q_{\frac{5m}{8}} = \mathbb{E}\{x\in X,  q_{\frac{m}{2}}< x\le q_{\frac{3m}{4}}\}$ and $q_{\frac{7m}{8}} = \mathbb{E}\{x\in X,  q_{\frac{3m}{4}}< x\le q_{m}\}$; 
\end{itemize}
and so forth. Each interval is assigned a $\mathrm{log_{2}}(m)$-bit Gray code sequence. Since RSSIs near the interval edges can cause bit disagreements, we drop the RSSIs within the guard bands $\big(q^{-}_{i}, q^{+}_{i}\big], 1\le i\le m-1$.  
After quantization, Alice and Bob obtain binary bit sequence, $K_A$ and $K_B$, respectively.
In summary, the quantizer converts analog measurements, $X$, into binary key bits using a quantization algorithm with  parameters including the quantization level $m$ and the guard band parameter $\alpha$, given as
\begin{align}
    K_u = \mathcal{Q}(X_u, m, \alpha),
\end{align}
where $u$ denotes the user.
The quantization parameters, $m$ and $\alpha$, are tunable. 

\begin{figure}[!t]
\centerline{\includegraphics[width=3.4in]{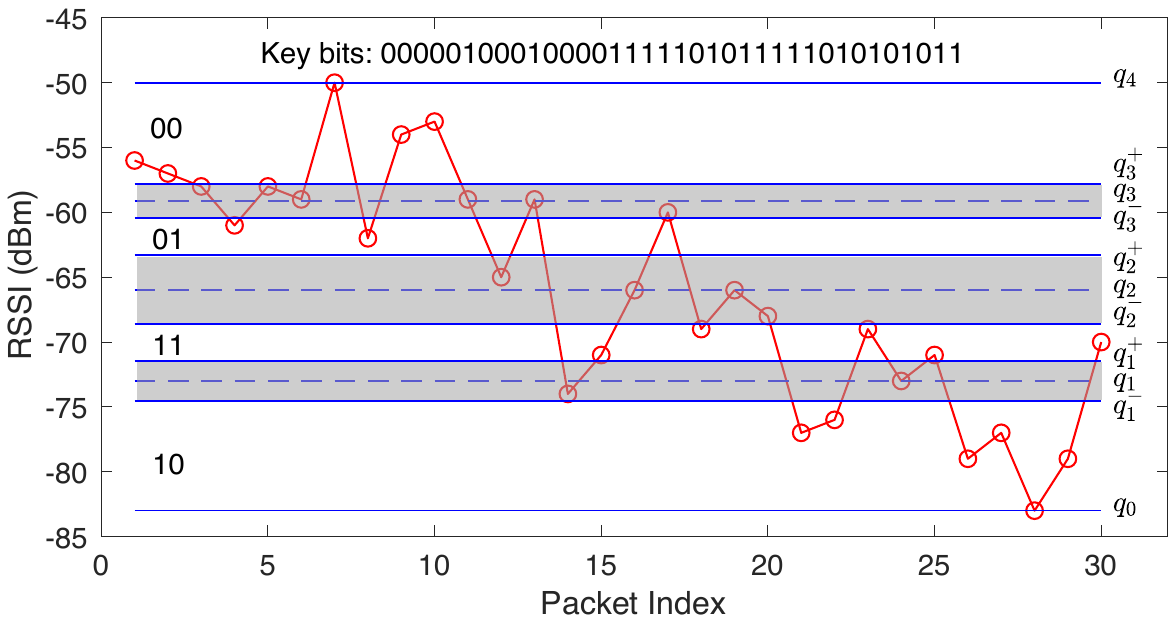}}
\caption{Multi-bit and lossy quantization with $m=4$ and $\alpha=0.25$. The shaded areas represent guard bands.}
\label{quan_example}
\end{figure}

\begin{figure}[!t]
 \centering
 \subfloat[]{\includegraphics[width=3.4in]{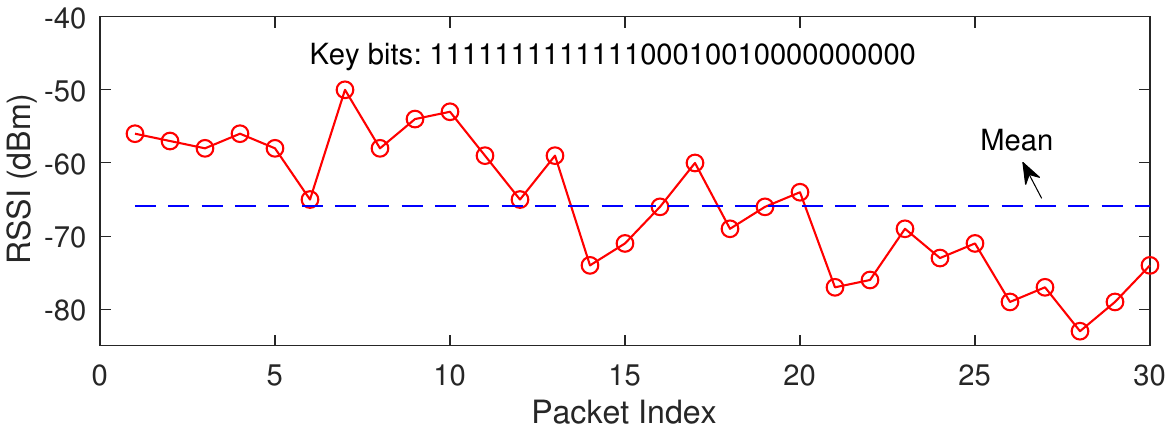}} \\
 \subfloat[]{\includegraphics[width=3.4in]{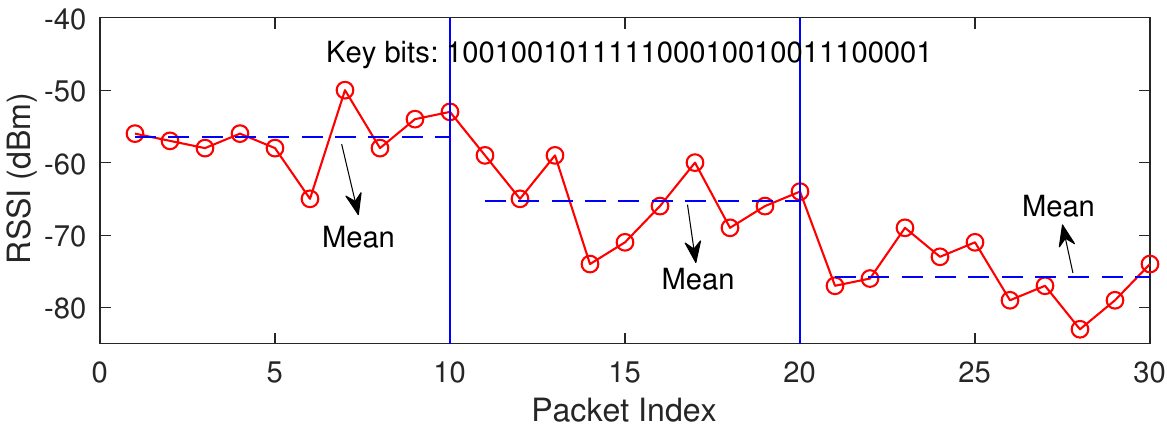}}
 \caption{Single-bit and lossless quantization with $m=2$ and $\alpha=0$. (a) Quantization block $L_{2}$=30. (b) Quantization block  $L_{2}$=10. }
\label{fig:quantization_block}
\end{figure}

An illustration of $m$ = 4-level quantization is shown in Fig.~\ref{quan_example}. 
The thresholds $q_n$ can be calculated as follows.
\begin{itemize}
	\item when $n=2$, $q_{2}=\mathbb{E}\{x\in X,  q_{0}\le x\le q_{4}\}$ and $s_{2}=\mathbb{S}\{x\in X,  q_{1}< x\le q_{3}\}$;
	\item when $n=1$, $q_{1}=\mathbb{E}\{x\in X,  q_{0}\le x\le q_{2}\}$ and $s_{1}=\mathbb{S}\{x\in X,  q_{0}\le x\le q_{2}\}$;
	\item when $n=3$, $q_{3}=\mathbb{E}\{x\in X,  q_{2}< x\le q_{4}\}$ and $s_{3}=\mathbb{S}\{x\in X,  q_{2}< x\le q_{4}\}$.
\end{itemize}
   After determining the quantization intervals, 2-bit Gray code sequences 10, 11, 01, and 00 are assigned to $Q_{1}$-$Q_{4}$, respectively. {When $m = 2$, it reduces to single-bit quantization, as illustrated in Fig.~\ref{fig:quantization_block}.}

{When $\alpha \neq 0$, there are guard bands. The larger the guard band parameter $\alpha$, the more RSSIs will be dropped, which results in lossy quantization, as shown in Fig.~\ref{quan_example}. In contrast, when $\alpha = 0$, no channel measurement is dropped, which is lossless quantization, as can be observed in Fig.~\ref{fig:quantization_block}.}

\subsubsection{Information Reconciliation}
\label{sec:Information_reconciliation}
Due to channel measurement errors caused by hardware imperfection, noise effects, and the delay between bidirectional sampling, Alice and Bob cannot obtain exactly the same measurements, which will result in key bit inconsistencies. Information reconciliation is used to address this issue, thereby ensuring that Alice and Bob generate the same key. Widely used information reconciliation schemes include  Cascade protocol and error correction codes (ECCs), e.g. BCH code and low-density parity-check code (LDPC). As shown in Fig. \ref{fig:keygen_steps}, Bob transmits the information necessary for error correction to Alice, with which
Alice corrects the disagreed key bits and obtains $K_{A^{'}}$.
Following key correction, key agreement check schemes, e.g., cyclic redundancy check (CRC), are adopted to confirm whether the two legitimate nodes generate the same key.
If the CRC check values are identical, Alice and Bob agree on the same key, i.e., $K_{A^{'}}=K_{B}$. 
Then Alice will send an information reconciliation character (IRC) to Bob to indicate whether the keys match. If the keys match, $\mathrm{IRC}=1$; otherwise, $\mathrm{IRC}=0$. Note that the roles of Alice and Bob are interchangeable in this process.

\subsubsection{Privacy Amplification}
Information exchanged through public channels during the information reconciliation stage could be leveraged by eavesdroppers. Privacy amplification aims to eliminate leaked information.

\subsection{Performance Metrics}
KGR evaluates the efficiency of key generation, defined as
\begin{align}
\mathrm{KGR} = \frac{N_{\mathrm{KEY}}}{N_{\mathrm{RSSI}}},
\end{align}
where $N_{\mathrm{RSSI}}$ is the number of RSSI measurements (packets) used to generate keys and $N_{\mathrm{KEY}}$ is the number of key bits obtained from these RSSIs.

KDR denotes the ratio of disagreed key bits between keys generated by Alice and Bob after quantization, defined as
\begin{align}
\mathrm{KDR} = \frac{\sum_{i=1}^{N_{\mathrm{KEY}}}\left|K_{A}(i)-K_{B}(i)\right|}{N_{\mathrm{KEY}}}.
\end{align}

To test the randomness of generated keys, we apply the widely-used NIST test suite \cite{NIST}. 
Each test will return a P-value. When the P-value is above 0.01, it indicates the sequence passes the applied test. Same as other works~\cite{TMC2,zhang2016experimental,IOTJ1}, we only used a subset of the 15 tests, as the rest of them require a long sequence which is currently not available. Because AES-128 encryption  is commonly used,  we perform the NIST test on each 128-bit key sequence. We define a new metric, NIST test failure ratio (NFR), which is the ratio of key sequences that do not pass the NIST test and computed by
\begin{align}
\mathrm{NFR} = \frac{N_{\mathrm{F}}}{N_{\mathrm{KS}}},
\end{align}
where $N_{\mathrm{KS}}$ is the  number of key sequences and $N_{\mathrm{F}}$ is the number of key sequences that fail to pass the NIST test.
 
\section{Problem Statement and System Overview}
\label{Problem Statement}
\subsection{Problem Statement}
Most of the existing key generation works focus on the design of channel probing that aims to exploit channel variations as much as possible. Under this direction, there have been research efforts exploring multiple antennas~\cite{zeng2010exploiting,li2022fast}, leveraging RISs~\cite{staat2021intelligent,li2021maximizing, lu2023joint} or using machine learning for enhancing measurements correlation~\cite{wei2022knew}. Nevertheless, quantization is largely overlooked but an improper quantization results in inefficient randomness extraction or even failure of key generation. 

This issue is more severe for long-distance transmission-based key generation because the channel between Alice and Bob could experience large variations. As illustrated in Fig.~\ref{fig:quantization_block}(a), quantizing a long sequence of RSSIs with large variations can result in long runs of consecutive 1s and 0s. 
In order to increase the randomness of generated keys, RSSI measurements can be partitioned into small quantization blocks and quantized block-wise. The length of a quantization block with quantization level $m$ is denoted by $L_{m}$. 
When we reduce the length of a quantization block from $L_m = 30$ in Fig.~\ref{fig:quantization_block}(a) to $L_m = 10$ in Fig.~\ref{fig:quantization_block}(b), the quality of the generated keys is improved.

\begin{figure}[!t]
 \centering
 \subfloat[]{\includegraphics[width=1.7in]{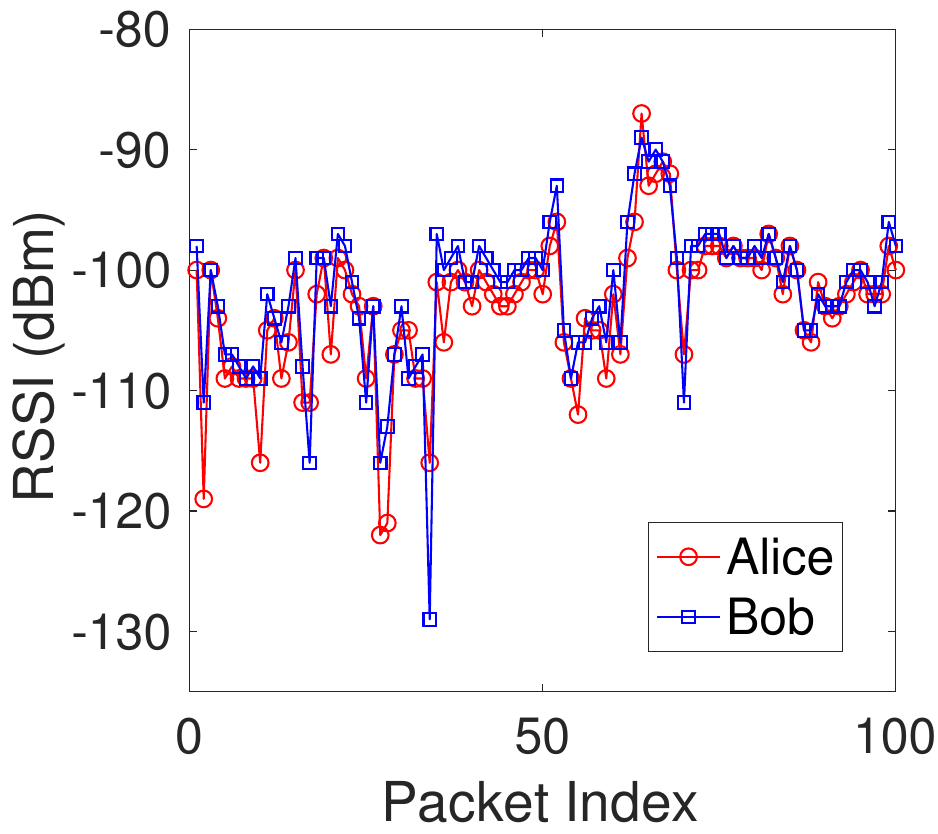}}
 \subfloat[]{\includegraphics[width=1.7in]{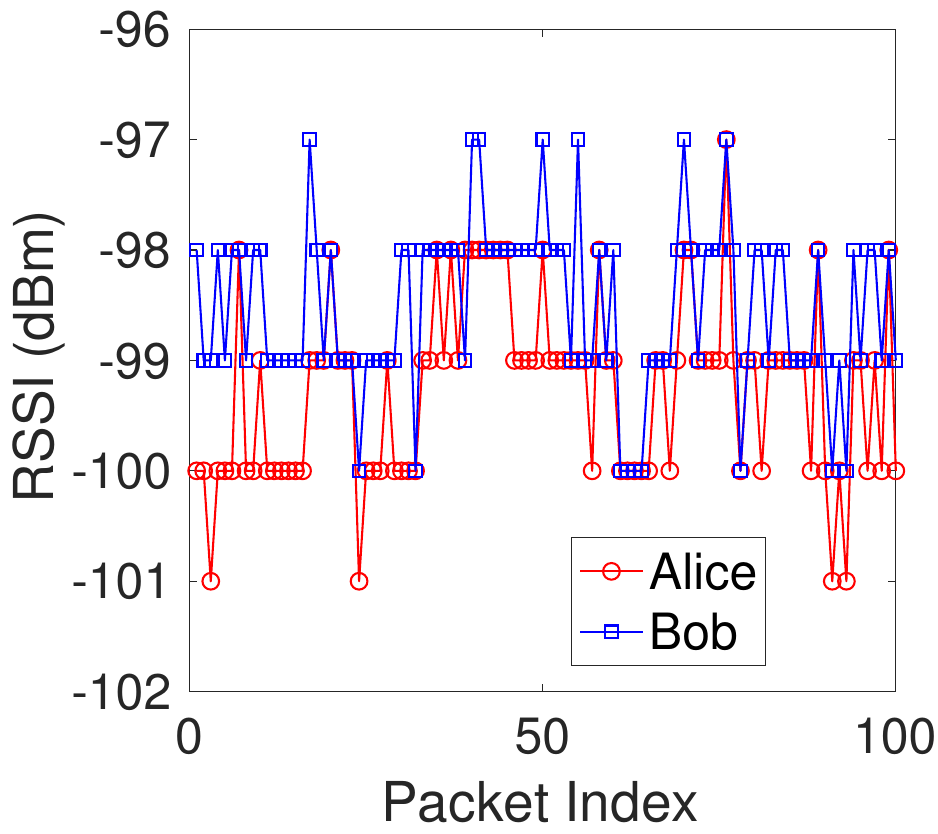}}
 \caption{RSSI measurements. (a) Sequence 1. (b) Sequence 2.}
\label{example_data}
\end{figure}

\begin{figure}[!t]
 \centering
 \subfloat[]{\includegraphics[width=3.4in]{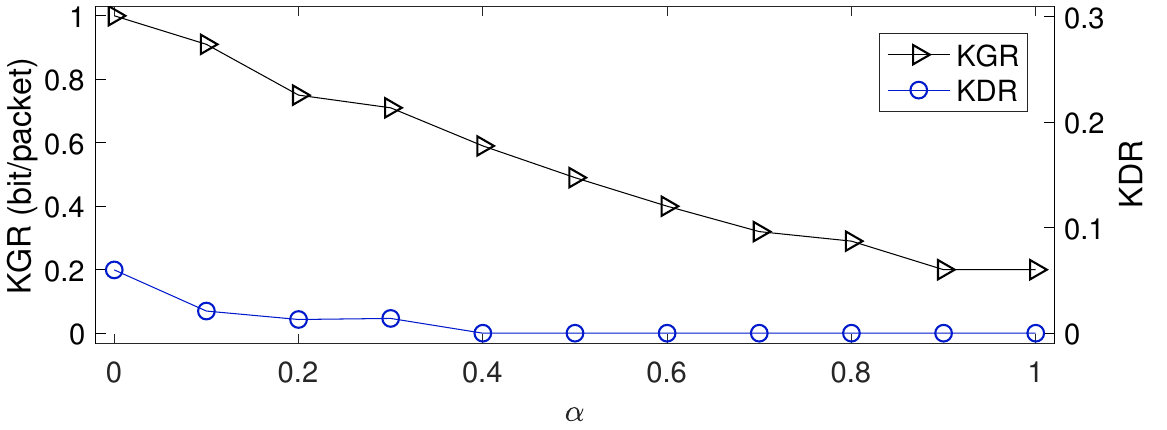}} \\
 \subfloat[]{\includegraphics[width=3.4in]{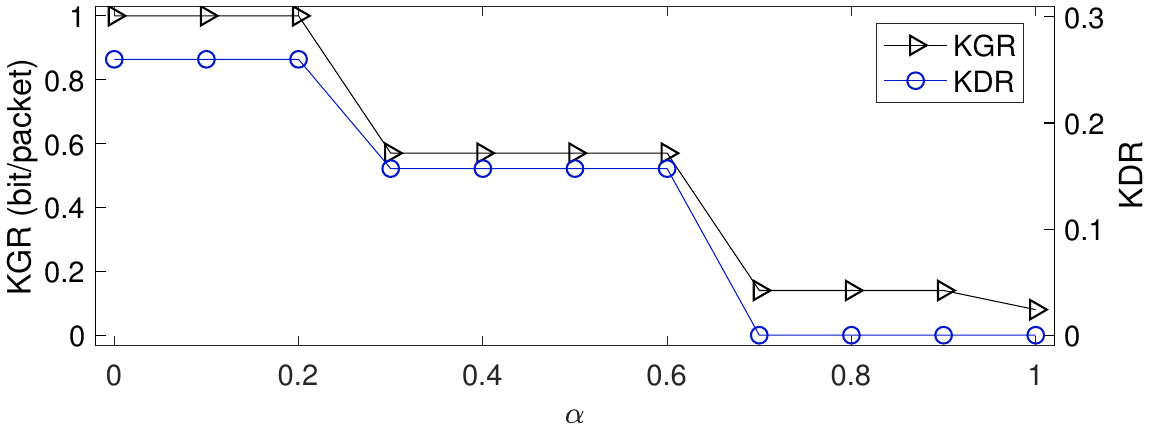}}
 \caption{Impact of $\alpha$ on KGR and KDR with $m=2$ and $L_2=100$. (a) Sequence 1. (b) Sequence 2.}
\label{example_data alpha}
\end{figure}

Different channel diversities may require different quantization parameters and achieve different KGRs. As shown in Fig.~\ref{example_data}, we collect two sequences of RSSI measurements using two LoRa devices (introduced in Section~\ref{sec:testbed}), where sequence 1 has larger variations than sequence 2. For ease of illustration, we adopt the 2-level quantization with $L_{2}=100$. Fig.~\ref{example_data alpha} shows the effect of $\alpha$ on KGR and KDR. We can see that a smaller $\alpha$ generally leads to a higher KGR, at the cost of a higher KDR. The KDR should be below a threshold that depends on the correction capability of the information reconciliation method used. In this paper, we adopt the BCH code for information reconciliation. A BCH code can correct up to 25\% mismatch \cite{zhang2016key}. To be on the safe side, we choose
a BCH code that can correct 20\% mismatch following~\cite{TVT1}. For sequence 1, the optimal $\alpha$  is 0 as it maximizes the KGR and keeps the KDR below 20\%. However, for sequence 2, the optimal $\alpha$ is 0.3 as a smaller $\alpha$ cannot satisfy the KDR requirement ($\leq 20\%$). Hence, sequence 1 and sequence 2 require different $\alpha$ to achieve their optimal KGRs.

In reality, channel measurements in different environments may have different channel diversities. Even in the same environment, channel diversity could vary over time. As a consequence, fixed
quantization parameters cannot adapt to the channel diversities
of different scenarios. Unfortunately,  KDR cannot be calculated in real applications because the RSSI measurements of Alice and Bob cannot be shared with each other; they are the keying materials that need to be kept secret.
How to choose the optimal quantization parameters remains a challenge but is extremely important as they determine the efficiency (KGR) or even success (KDR) of the key generation. 

\subsection{System Overview}
To address this challenge, we propose
a novel adaptive quantization scheme that can dynamically adjust quantization parameters as per local RSSI measurements.
The proposed AQ-KG system consists of two parts: offline training for adaptive quantization and online deployment of AQ-KG protocol. 
\subsubsection{Offline Training}
The core idea of this paper is to first obtain optimal quantization parameters for the collected RSSI dataset, and then develop linear regression models to learn the mapping between the randomness of RSSIs, which is  quantified by Lempel-Ziv complexity (LZ76), and the optimal quantization parameters.  This process is conducted in the offline training phase and is detailed in Section \ref{Adaptive Quantization}. 

\subsubsection{Online Deployment}
The pretrained linear regression models can be used to adaptively determine quantization parameters for new RSSI measurements in real-time.  Based on the pre-trained adaptive quantization scheme,  a complete AQ-KG protocol is designed in Section \ref{sec:keygen}.  

\subsection{Measurement Testbed}\label{sec:testbed}
\begin{figure}[!t]
 \centering
 \subfloat[]{\includegraphics[height=1.28in]{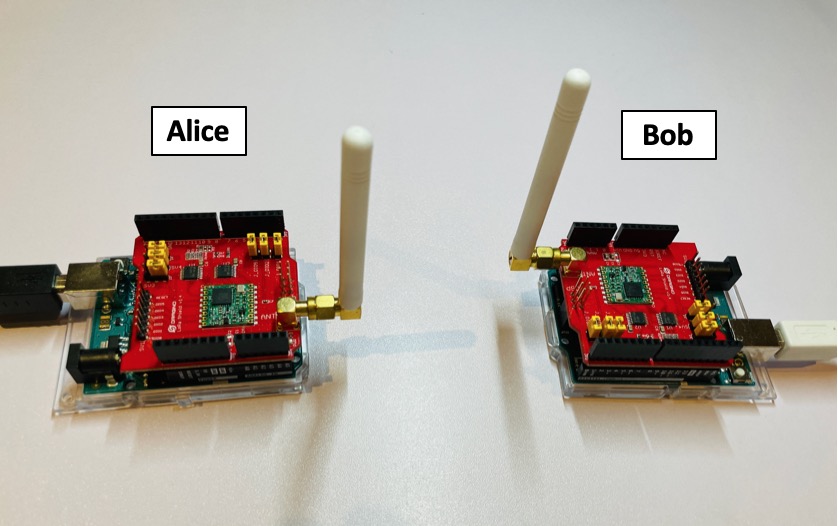}}
 \subfloat[]{\includegraphics[height=1.28in]{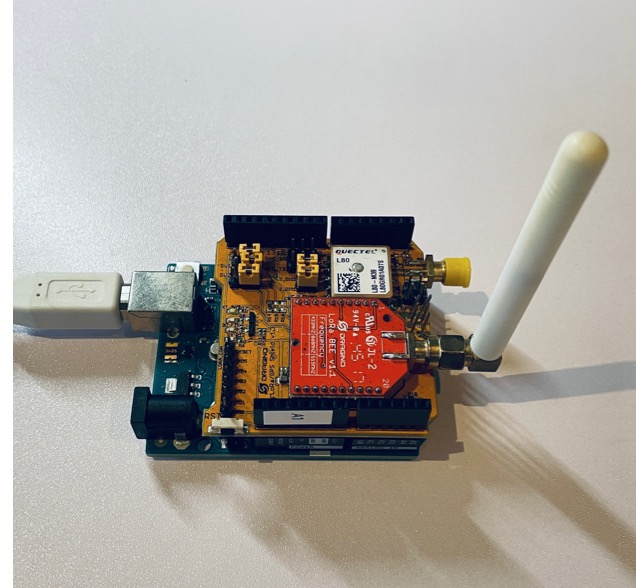}}
 \caption{Hardware platform. (a) Dragino LoRa Shield v1.4. (b) Dragino LoRa/GPS Shield v1.3.}
\label{device}
\end{figure}

We established a testbed using Dragino LoRa Shield v1.4 (Fig.~\ref{device}(a)) as well as Dragino LoRa/GPS Shield v1.3 (Fig.~\ref{device}(b)). As further elaborated in Table~\ref{tab1}, we considered two types of setup: (i) identical LoRa modules, i.e., both Alice and Bob were using Dragino LoRa Shield v1.4; and (ii) different LoRa modules, i.e., Alice was using Dragino LoRa Shield v1.4  and Bob was using Dragino LoRa/GPS Shield v1.3. Without otherwise specified, identical modules were used.

All the LoRa modules were configured with the same parameters including carrier frequency of 868.1 MHz, bandwidth of 125 kHz, transmit power of 13 dBm, and spreading factor of 7.  Alice and Bob perform channel probing as described in Section~\ref{sec:Channel_Probing1}. 
The collected RSSIs are transferred to the laptop through a serial port for further processing using Python.
  
\section{Offline Training of Adaptive Quantization}
\label{Adaptive Quantization}

\subsection{Design Philosophy}
\begin{figure*}[!t]
\centerline{\includegraphics[scale=0.53]{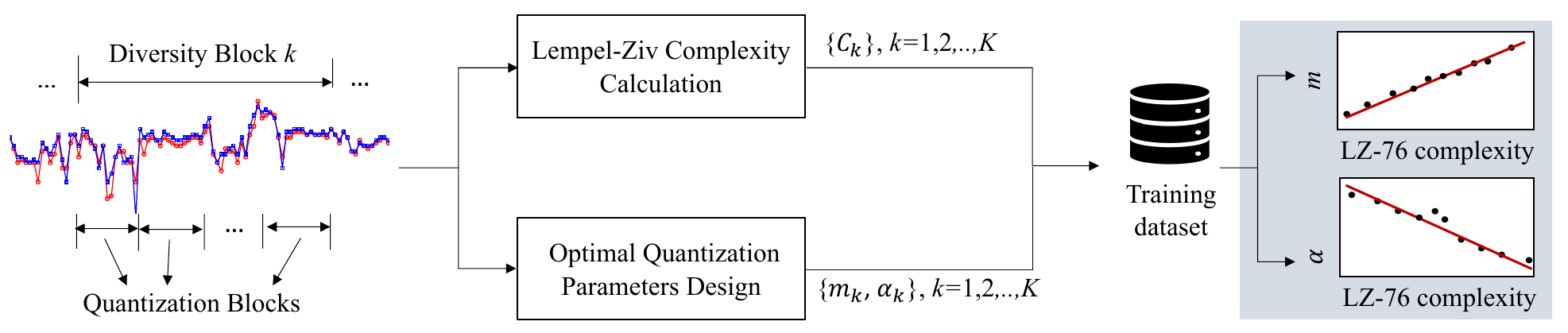}}\caption{Block diagram of the offline training for adaptive quantization. The gray box represents the training of the linear regression models, which will be used to determine the quantization parameters.}
\label{workflow_training}
\end{figure*}

The randomness of RSSI measurements varies over time, and thus quantization parameters need to be adjusted accordingly. To this end, we propose to partition the collected RSSIs into blocks and select appropriate quantization parameters separately for each block. The block length should be small so that the channel diversity within each block is stable. In this paper, we refer to such blocks as diversity blocks, where quantization parameters $m$ and $\alpha$ remain unchanged within a diversity block but may differ across diversity blocks. In each diversity block, the optimal quantization parameters should maximize the KGR while keeping the KDR below 20\%. The length of a diversity block is denoted by $L_{\mathrm{D}}$. As will be shown in Section~\ref{sec:block_dertermination}, different quantization levels require different quantization block lengths. To determine the quantization level and perform quantization in a diversity block, we have $L_{\mathrm{D}}\ge L_{m}, \forall m\in \mathcal{M}$, where $\mathcal{M}$ is the set of $m$.
A diversity block may contain multiple quantization blocks. An illustration can be found in Fig.~\ref{workflow_training}.

Fig. \ref{workflow_training} shows the offline training process. 
The optimal quantization parameters and LZ76 complexity are calculated for each diversity block. The LZ76 complexities and their corresponding optimal quantization parameters constitute the training dataset. Based on the training dataset, linear regression models are developed to learn the relationship between the randomness of RSSIs (LZ76 complexity) and the optimal quantization parameters. 

\subsection{RSSI Dataset}
\label{sec:RSSI_Dataset}
In order to ensure the generalization of the RSSI dataset, we have specifically designed different measurement scenarios.

Alice was always located on the 4th floor of a residential building in York, U.K. 
Regarding Bob, we considered different locations (indoor and outdoor) and moving patterns (stationary and moving).
Therefore, there are four scenarios in total.
\begin{itemize}
\item Scenario A: Outdoor and moving. Bob was moving around the residential building at a walking speed.
\item Scenario B: Outdoor and stationary. Bob was placed in the city center of York, about 500 meters away from Alice.
\item Scenario C: Indoor and moving. Bob was moving inside the residential building at a walking speed.
\item Scenario D: Indoor and stationary.  Bob was placed on the 4th floor of the residential building, about 20 meters away from Alice. 
\end{itemize}
When both Alice and Bob are stationary, the randomness required for key generation relies on environmental changes, e.g., moving pedestrians and vehicles. When Bob is moving, the wireless channel is changing more dynamically.

\begin{figure}[!t]
 \centering
 \subfloat[]{\includegraphics[width=1.675in]{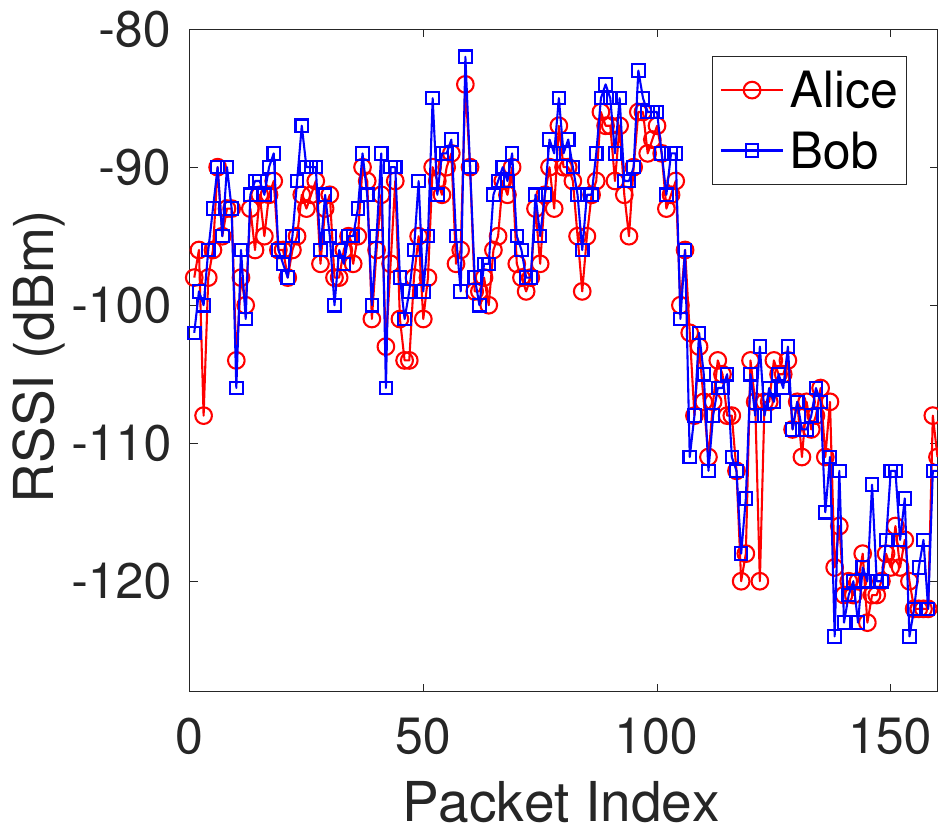}} \hspace{0.05in}
 \subfloat[]{\includegraphics[width=1.675in]{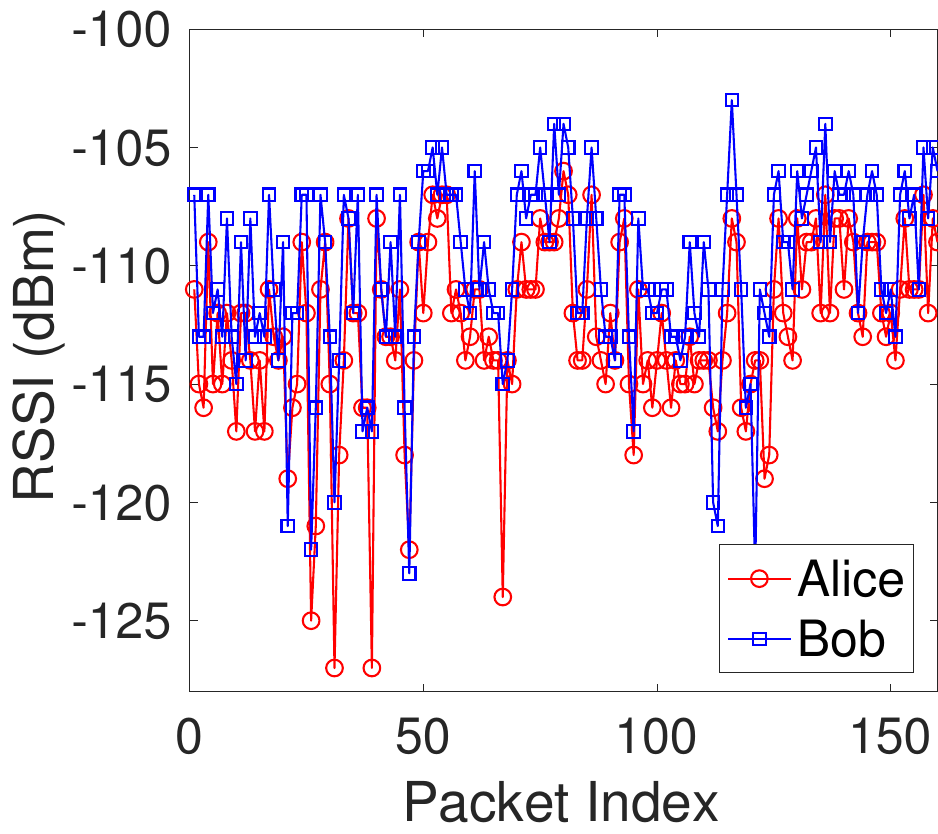}}
 \quad 
  \subfloat[]{\includegraphics[width=1.675in]{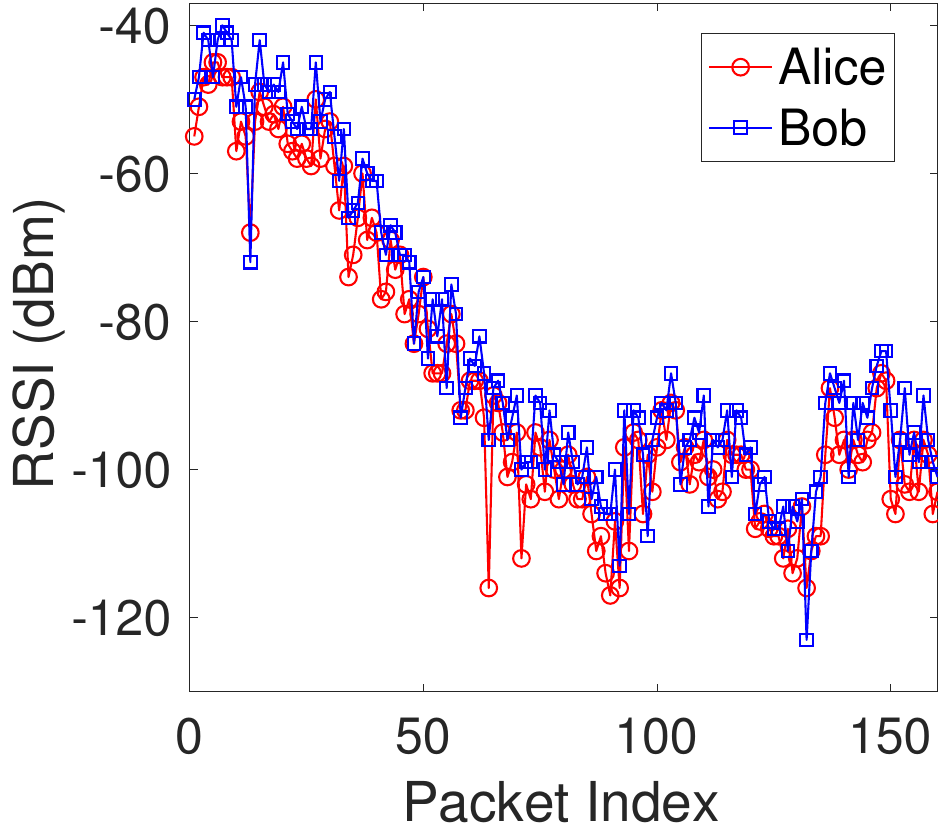}} \hspace{0.05in}
   \subfloat[]{\includegraphics[width=1.675in]{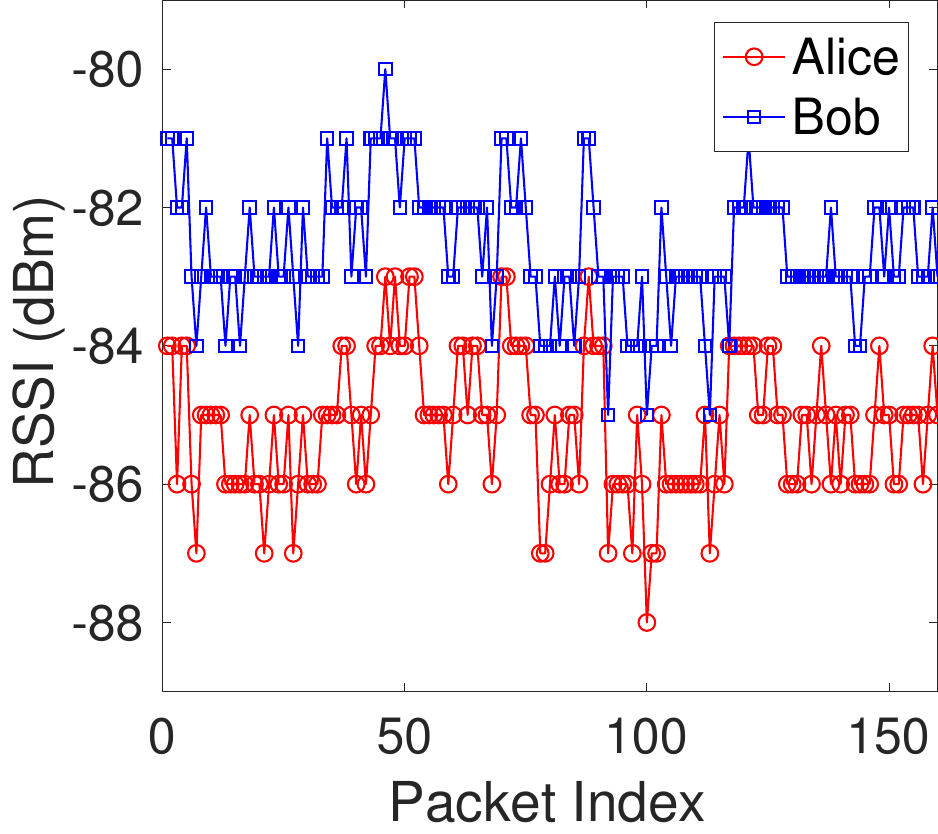}}
\caption{RSSI measurements in different scenarios. (a) Scenario A. (b) Scenario B. (c) Scenario C. (d) Scenario D.}
\label{Rssi}
\end{figure}

We performed RSSI measurements in the aforementioned four scenarios and collected about 140,000 RSSIs at different times on different days.
Fig. \ref{Rssi} shows part of the RSSI sequences collected in Scenario A-D.  We can see that RSSI measurements in moving scenarios have larger variations. 

\subsection{Determination of Quantization Block and Diversity Block}
\label{sec:block_dertermination}
The lengths of the quantization block and diversity block, $L_m$ and $L_{\mathrm{D}}$, are essential as they will affect the randomness of the keys generated. The values of $L_{m}$ and $L_{\mathrm{D}}$ are determined as follows. 

\begin{figure}[!t]
 \centering
 \subfloat[]{\includegraphics[width=3.4 in]{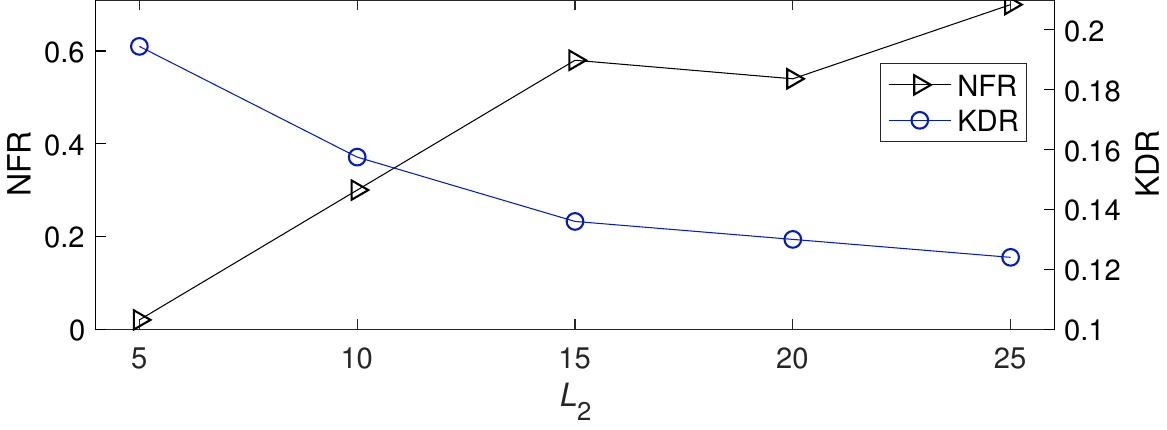}} \\
 \subfloat[]{\includegraphics[width=3.4 in]{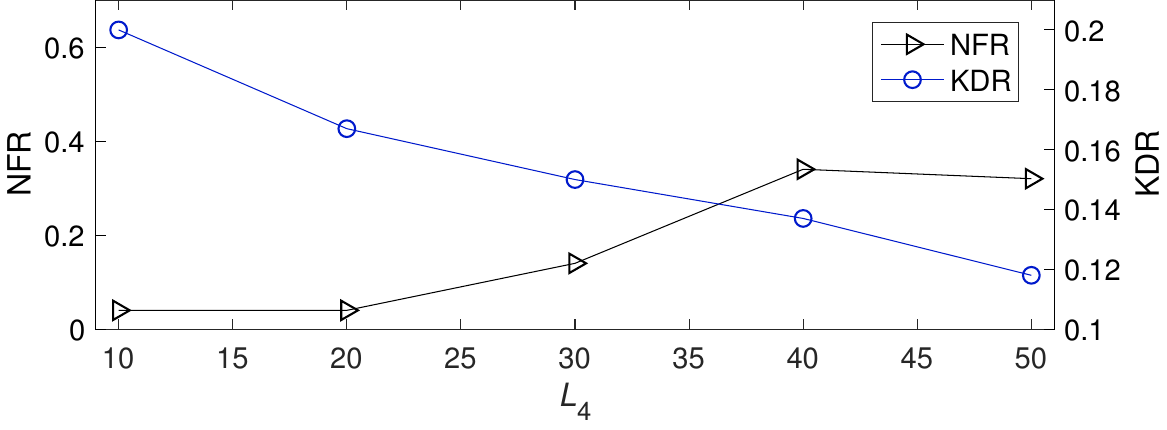}} \\
   \subfloat[]{\includegraphics[width=3.4 in]{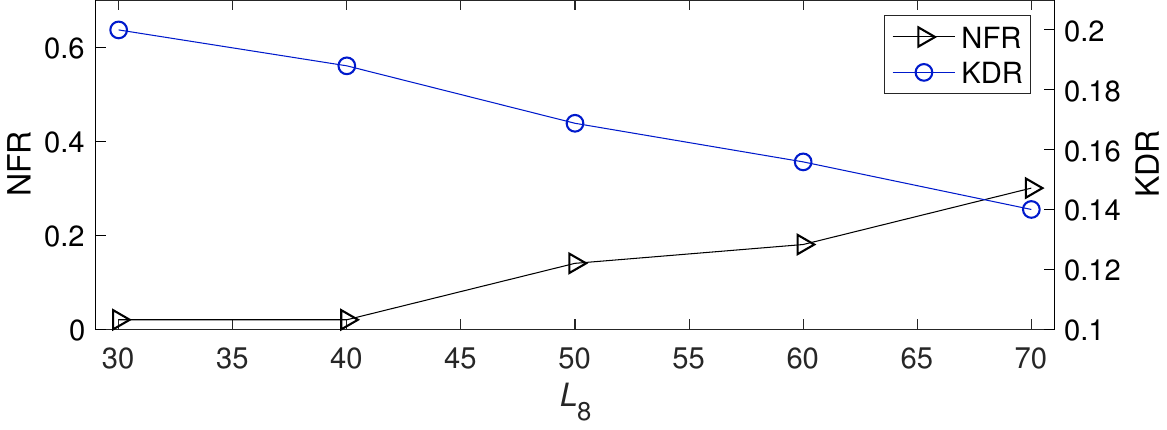}}
\caption{Impact of the quantization block length. (a) $L_2$. (b) $L_4$. (c) $L_8$.}
\label{blocklength}
\end{figure}

To obtain the optimal $L_{m}$, we performed  multi-level quantization on RSSI measurements collected from Scenario A and Scenario C as RSSI measurements in dynamic scenarios have larger variations. We generated 50 128-bit key sequences using different sets of quantization parameters and conducted the NIST test on the generated key sequences.
Fig. \ref{blocklength} shows the impact of $L_{m}$ on NFR and KDR. 
To maximize the KGR and keep the KDR below 20\%, we set $\alpha=0.1,0.35,0.5$ for $m=2,4,8$, respectively. 
In general, NFR increases with $L_{m}$ and KDR decreases with $L_{m}$. Hence, we select the largest $L_{m}$ that keeps the NFR low. Accordingly, we set $L_{2}=5$, $L_{4}=20$ and $L_{8}=40$. We observe that $m=16$ cannot keep the KDR below 20\% while achieving a low NFR. Hence, we consider $m\le 8$ in this paper. 

In order to perform quantization within a diversity block, we have  $L_{\mathrm{D}}\ge \mathrm{max}\{L_{2}, L_{4}, L_{8}\}$.  Meanwhile,  $L_{\mathrm{D}}$ should be small so that the  channel diversity is stable in a diversity block.  Thus, we set $L_{\mathrm{D}}=\mathrm{max}\{L_{2}, L_{4}, L_{8}\}=40$.

\subsection{Optimal Quantization Parameters Design}
In this subsection, we introduce how to obtain optimal quantization parameters for the collected RSSI dataset. The quantization parameters include quantization level $m$ and guard band parameter $\alpha$. 

The RSSI dataset is first divided into multiple diversity blocks. In each diversity block, we iteratively search for the optimal combination of $m$ and $\alpha$ that maximizes the KGR while keeping the KDR below 20\%. Since a larger $\alpha$ drops more RSSIs, for a given quantization level, the optimal $\alpha$ is the smallest one that keeps the KDR below 20\%. The details of obtaining optimal quantization parameters are given in Algorithm 1.  Note that this algorithm can only be executed in the  offline training phase when both Alice's and Bob's RSSI measurements are available, which is not feasible in practice. 
\begin{algorithm}
\label{algorithm1}
	\renewcommand{\algorithmicrequire}{\textbf{Input:}}
	\renewcommand{\algorithmicensure}{\textbf{Output:}}
	\caption{Calculate optimal quantization parameters}
	\label{alg1}
	\begin{algorithmic}[1]
	\REQUIRE $X_{A}$,  $X_{B}$ \ \% RSSI measurements of Alice and Bob
		\STATE Partition $X_{A}$ and $X_{B}$ into $K$ diversity blocks, respectively
		\FOR{$k$ := 1 to $K$}
		\FOR{$m$ := 2, 4, 8}
		\STATE $\alpha_{k}^{m} \gets 0$ \ \% Initialize guard band parameter 
		\STATE Partition diversity block $k$ into $L_{\mathrm{D}}$/$L_{m}$ quantization blocks
		\STATE Perform quantization in each quantization block with $m$ and $\alpha_{k}^{m}$
		\STATE Update $\mathrm{KDR}_{k}^{m}$ \% Calculate KDR for diversity block $k$
		\STATE Update $\mathrm{KGR}_{k}^{m}$ \% Calculate KGR for diversity block $k$
		\WHILE{$\mathrm{KDR}_{k}^{m}$ $>$ 20\%} 
		\STATE $\alpha_{k}^{m} \gets \alpha_{k}^{m}+0.01$
		\STATE Update $\mathrm{KDR}_{k}^{m}$
		\STATE Update $\mathrm{KGR}_{k}^{m}$
		\ENDWHILE
		\ENDFOR
		\STATE $m_{k}\gets$ $\underset{m}{\mathop{\arg\max}}$  $\mathrm{KDR}_{k}^{m}$
		\STATE $\alpha_{k} \gets \alpha_{k}^{m_{k}}$ 
		\ENDFOR
	\ENSURE  $m_{k}, \alpha_{k}, k=1,2,\dots,K$
	\end{algorithmic}  
\end{algorithm}

\subsection{Lempel-Ziv Complexity Calculation}
We adopt LZ76 \cite{TIT1} to measure the randomness of RSSIs since it is easy to calculate from experimental data. For a finite sequence $X$ with length $l_{X}$, $X(i,j)$ is a substring of $X$  that begins at index $i$ and ends at index $j$ when $i\le j$; while $X(i,j)$ is an empty string when $i>j$.  $X$ is considered to be reproducible from its substring $X(1,j)$ if $X(j+1,l_{X})$ is a substring of $X(1,l_{X}-1)$. This reproducibility is denoted by $X(1,j)\to X$. For example,  $101 \to 1010101$ as $0101$ is a substring of $101010$. A substring $X(i,j)$ is said to be exhaustive if $X(1,i-1)\to X(1,j-1)$ but $X(1,i-1)\nrightarrow X(1,j)$. For example, if $X=10011$, $X(3,4)=01$ is exhaustive as $X(1,2)\to X(1,3)$ but $X(1,2)\nrightarrow X(1,4)$; $X(3,5)=011$ is not exhaustive as $X(1,2)\nrightarrow X(1,4)$. Said differently, $X(i,j)$ is exhaustive if it is the shortest string that starts at index $i$ and is not a substring of $X(1,j-1)$. 

Decompose $X$ into $t$ components as follows:
\begin{align}
D(X) = X(1,c_{1})X(c_{1}+1,c_{2})\dots X(c_{t-1}+1,c_{t}),
\end{align}
where $c_{1}=1$ and $c_{t}=l_{X}$. Notice that the first component is always $X(1,1)$. The decomposition $D(X)$ is considered as exhaustive when all the components except the last one are exhaustive. The last component can be either exhaustive or not exhaustive. The LZ76 complexity $C_{LZ76}(X)$ is defined as the number of components when $D(X)$ is exhaustive. Generally a larger LZ76 complexity indicates a higher randomness.

For ease of understanding, we give three instances. For a binary string $X=01001110$, the exhaustive decomposition is $D(X)=0\cdot1\cdot00\cdot11\cdot10$ and $C_{LZ76}(X)=5$. LZ76 can also be used to analyze natural text \cite{PLOS1}. Assuming a text string $X=\mathrm{standards}$, the exhaustive decomposition is $D(X)=\mathrm{s\cdot t\cdot a\cdot n\cdot d\cdot ar\cdot ds}$ and $C_{LZ76}(X)=7$. The same partition rules are followed for RSSI measurements \cite{TMC1,wei2011adaptive}. Assuming a RSSI sequence $X=\{-50,-55,-50,-50,-48,-50,-55,-50\}$, we have $D(X)=-50\cdot-55\cdot-50-50\cdot-48\cdot-50-55-50$ and $C_{LZ76}(X)=5$.

In this paper, we adopt normalized LZ76 complexity to  evaluate the randomness of RSSI measurements in a diversity block, which is given by
\begin{align}
C(X_{k}) = \frac{C_{LZ76}(X_{k})}{L_{\mathrm{D}}},
\end{align}
where $X_{k}$ is the sequence of RSSIs in diversity block $k$. 

\subsection{Training of Linear Regression Models}
\label{Adaptive Quantization Section}

After calculating the optimal quantization parameters and the normalized LZ76 complexity for each diversity block, we obtain the training dataset. Then we train linear regression models to  learn the mapping between the quantization parameters ($m$ and $\alpha$) and the normalized LZ76 complexity.
 
\begin{figure}[!t]
\centerline{\includegraphics[width=3.4in]{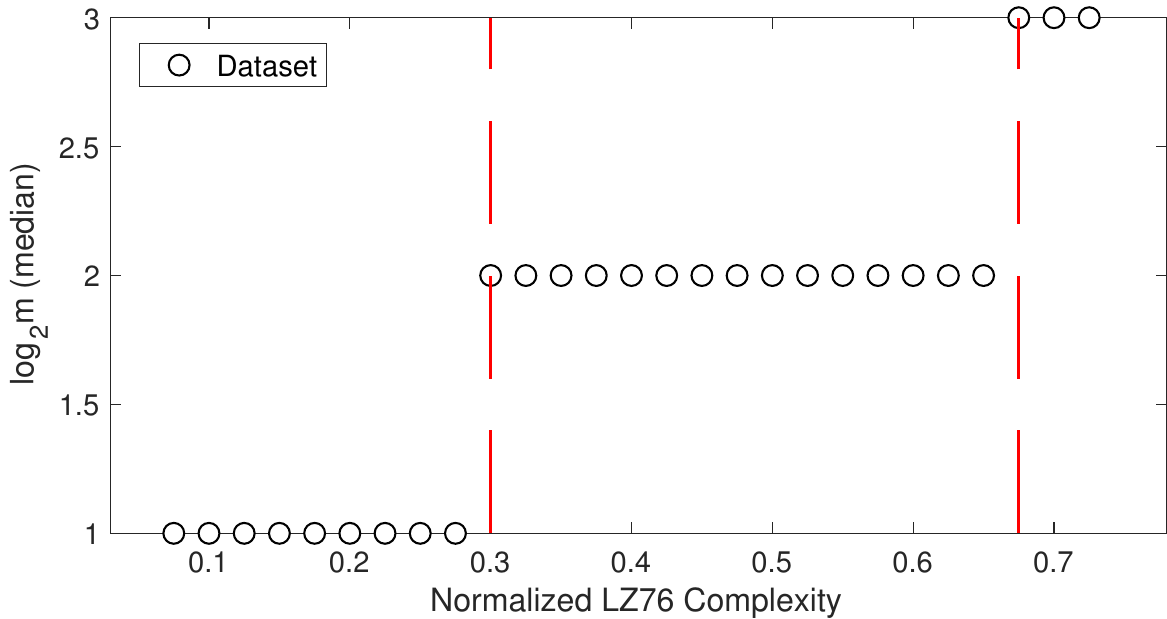}}
\caption{Normalized LZ76 complexity v.s. the median of quantization bits.}
\label{quan_type}
\end{figure}

 We first learn how to choose $m$ for a given normalized LZ76 complexity.
As multiple diversity blocks may correspond to the same normalized LZ76 complexity, each normalized LZ76 complexity in the training dataset may correspond to multiple different values of $m$. For example, the set of $m$ corresponding to normalized LZ76 complexity 0.675 is $\{8, 4, 4, 8, 8, 4, 8, 2, 8, 8, 8, 8, 4, 8, 8\}$.  
In Fig. \ref{quan_type}, we show the median of $\mathrm{log_{2}}(m)$ versus the normalized LZ76 complexity.  When $C(X_{k})< 0.3$, we set $m_{k}=2$; when  $0.3\le C(X_{k})< 0.675$,  we set $m_{k}=4$; otherwise, we set $m_{k}=8$. We choose the median rather than the mean to mitigate the deviations caused by extremely large or small values. It is observed that higher LZ76 complexity generally leads to a higher optimal quantization level because larger channel variations can yield more key bits.

\begin{figure}[!t]
 \centering
 \subfloat[]{\includegraphics[width=3.4 in]{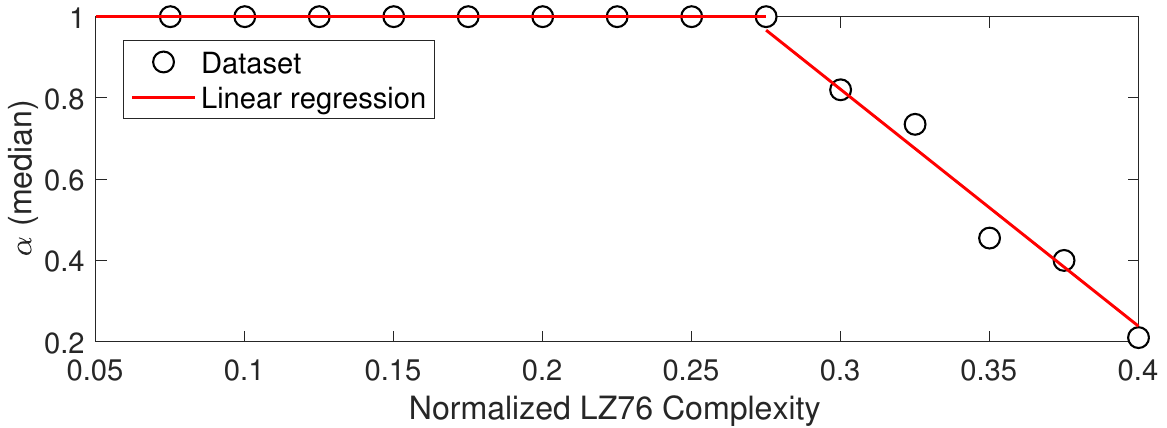}}\\
 \subfloat[]{\includegraphics[width=3.4 in]{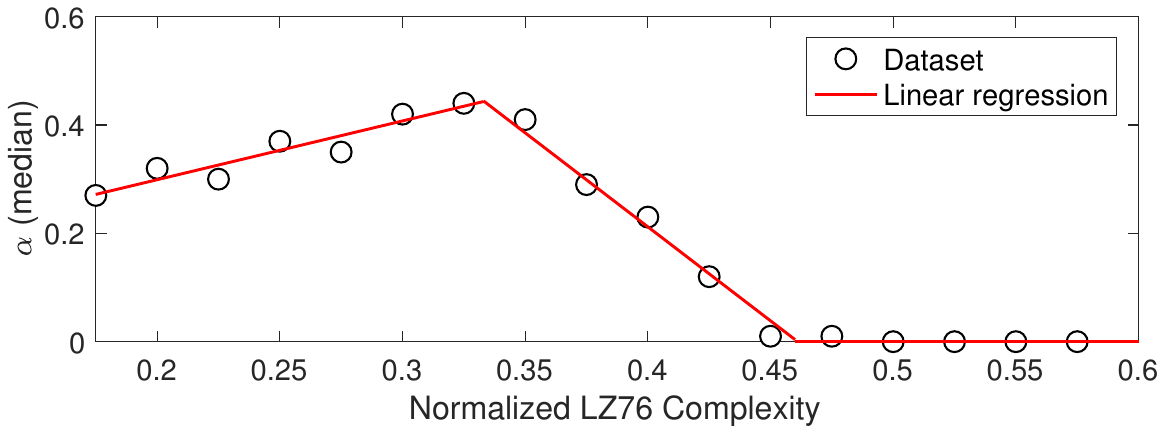}}\\
   \subfloat[]{\includegraphics[width=3.4 in]{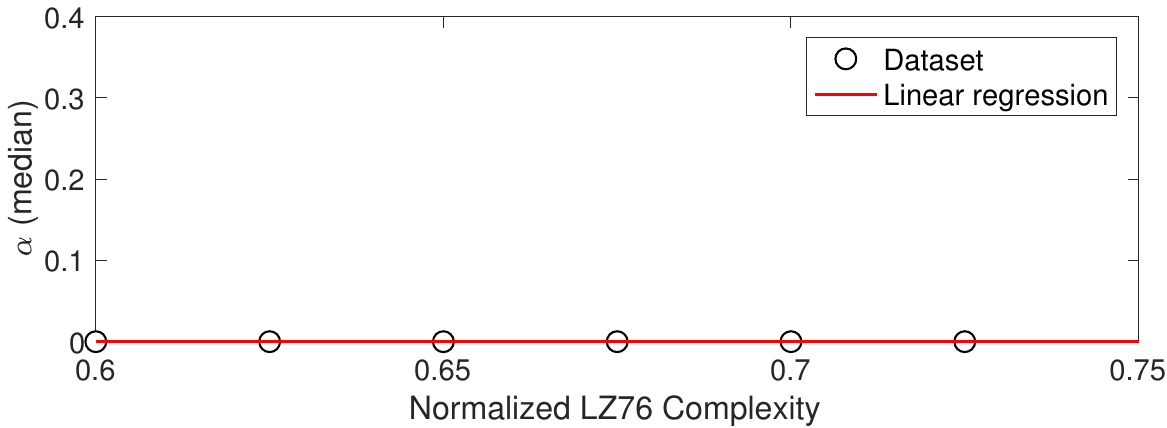}}
\caption{Normalized LZ76 complexity v.s. median of $\alpha$. (a) $m=2$. (b) $m=4$. (c) $m=8$.}
\label{quan_alpha}
\end{figure}

After determining  $m_{k}$, we select an appropriate $\alpha_{k}$ for the given $m_{k}$. To this end, we group all the diversity blocks into $\mathcal{S}_{2}$, $\mathcal{S}_{4}$ and $\mathcal{S}_{8}$, where $\mathcal{S}_{m}$ denotes the set of diversity blocks with the same optimal quantization level $m$. For each $\mathcal{S}_{m}$, we calculate the median of $\alpha$ corresponding to each normalized LZ76 complexity. Fig. \ref{quan_alpha} shows the normalized LZ76 complexity versus the median of $\alpha$ for $m=2$, $m=4$ and $m=8$, respectively. We can observe that there is a clear linear relationship between the normalized LZ76 complexity and the optimal
quantization level. This motivates us to train a linear regression model for each quantization level. For a RSSI sequence $X_{k}$, we can get the following models:
\begin{itemize}
    \item When $m_{k}=2$ 
\begin{align}
\alpha_{k} \!=\! 
 \left\{\!
\begin{array}{ll}
  1, & {C(X_{k}) \le 0.275}
\\
 -5.83C(X_{k}) + 2.57, &  {C(X_{k})> 0.275}
\end{array} \right.
\label{alpha_m2}
\end{align}

\item When $m_{k}=4$
\begin{align}
\alpha_{k} \!=\! 
 \left\{\!
\begin{array}{ll}
 1.085C(X_{k}) \!-\! 0.082, & {C(X_{k})\le 0.33}
\\
 -3.47C(X_{k}) \!+\! 1.6, &  {0.33\!<\!C(X_{k})\le 0.46}
 \\
 0, & {C(X_{k}) > 0.46}
\end{array} \right.
\label{alpha_m4}
\end{align}

\item When $m_{k}=8$
\begin{align}
\alpha_{k} \!=\! 0.
\label{alpha_m8}
\end{align}
\end{itemize}

\begin{figure}[!t]
\centerline{\includegraphics[width=3.4in]{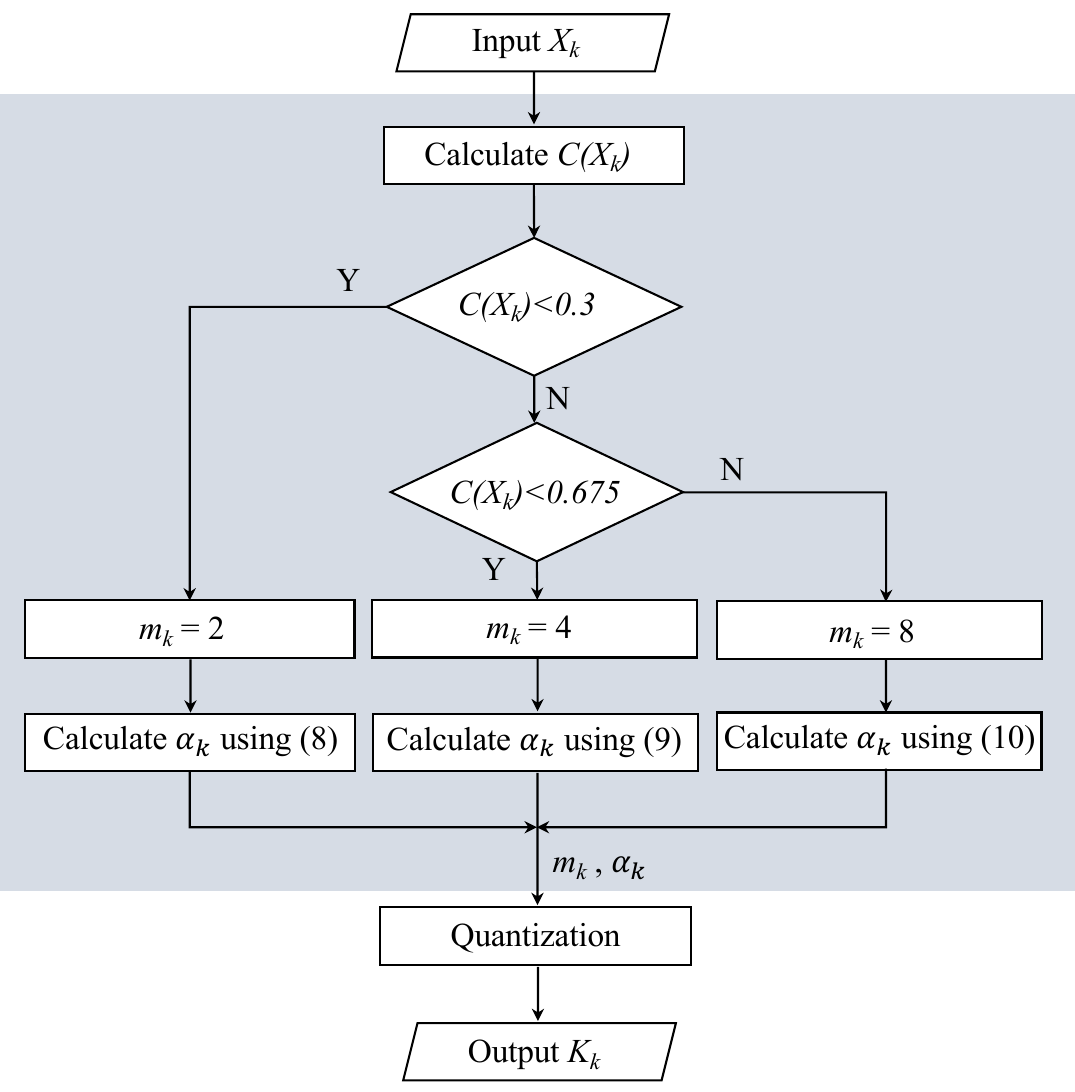}}
\caption{Flow chart of determining quantization parameters.}
\label{AQ}
\end{figure}

The flow chart of the proposed adaptive quantization scheme is shown in Fig. \ref{AQ}. After offline training, our proposed scheme is able to adjust quantization parameters for new RSSI measurements. It is worth mentioning that in Fig.~\ref{quan_type} and Fig.~\ref{quan_alpha}, the normalized LZ76 complexities are calculated using the RSSI values collected by Alice. Nevertheless, thanks to the reciprocity of wireless channels, we find that the same linear regression models can be obtained using the RSSI values collected by Bob. Therefore, either Alice and Bob can determine the quantization parameters using the proposed adaptive quantization scheme and share the quantization parameters with the other legitimate node. For the sake of illustration, in the following, we assume that Alice determines the quantization parameters and then shares them with Bob. Note that both the dataset collection and the model training are performed offline and only need to be done once. During the online deployment, the low-complexity linear regression models are computationally efficient.

\section{Real-Time Adaptive Quantization-Based Key Generation}\label{sec:keygen}
\begin{figure}[!t]
\centerline{\includegraphics[width=3.4in]{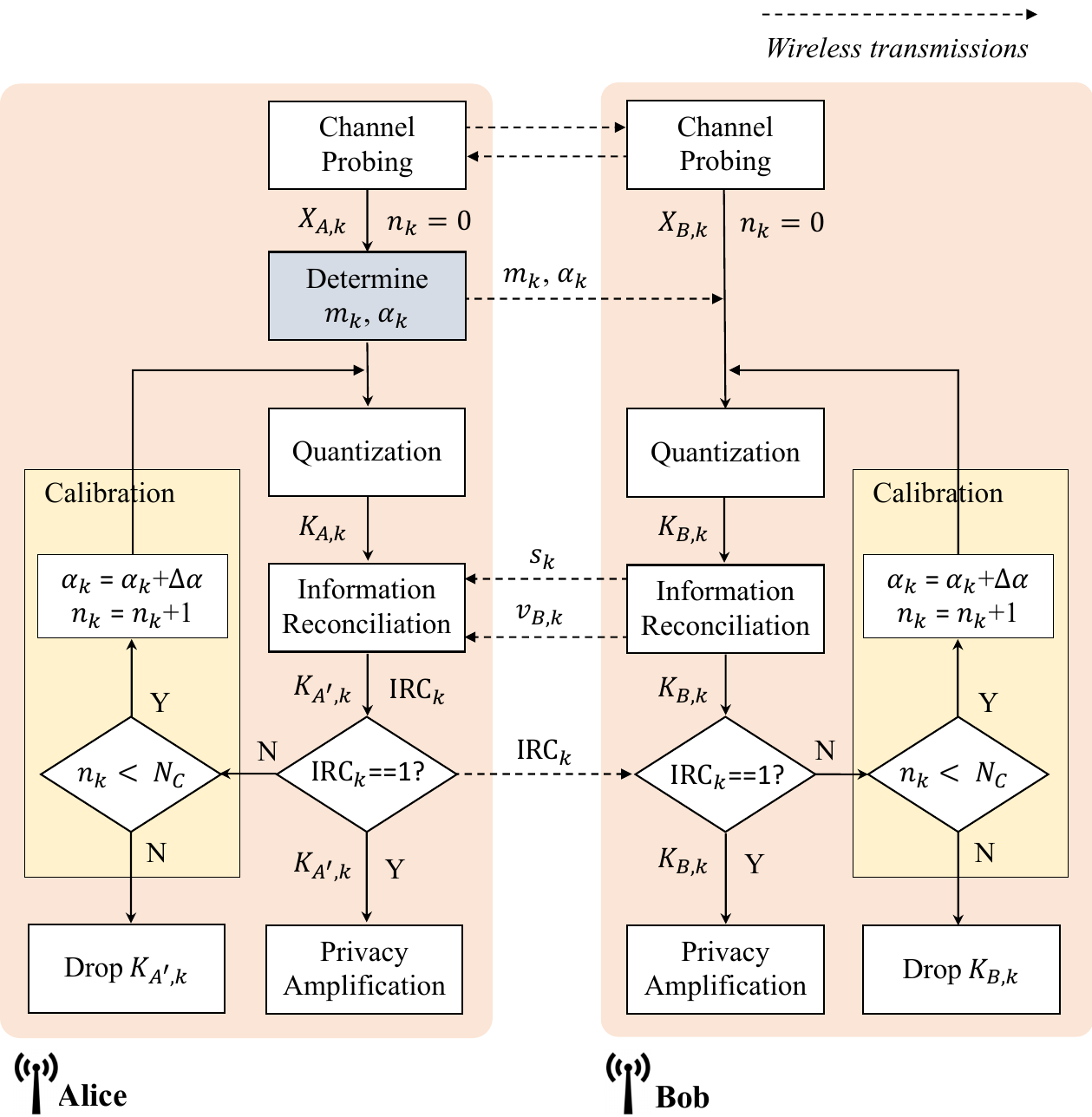}}
\caption{Workflow of the proposed AQ-KG protocol.}
\label{workflow}
\end{figure}

In this section, we present the steps of the proposed AQ-KG protocol for online deployment. The workflow of the proposed AQ-KG protocol is shown in Fig. \ref{workflow}.
The new modules compared to the protocol shown in Fig.~\ref{fig:keygen_steps} are highlighted in different colors. The details of each step are introduced as follows.

\subsection{Channel Probing}
\label{sec:Channel_Probing}
Two legitimate nodes, Alice and Bob, perform bidirectional channel measurements and record the RSSI value of each measurement. 
 Due to the long transmission distances or severe penetration loss, there could be a loss of packets. Hence, the indices of lost packets should be shared between Alice and Bob to ensure that the measurements match.

\subsection{Adaptive Quantization}
After Alice and Bob collect enough RSSI measurements, they partition the measurements into diversity blocks.
In the diversity block $k$, Alice determines $m_{k}$ and $\alpha_{k}$ and sends them to Bob.
The indices of RSSI measurements fall in guard bands are shared between Alice and Bob to keep the RSSI measurements with common indices.
Then Alice and Bob perform quantization on their RSSI sequences $X_{A,k}$ and $X_{B,k}$, and  obtain key sequences $K_{A,k}$ and $K_{B,k}$, respectively.

\subsection{Information Reconciliation}
\subsubsection{Correction and Agreement Check}
The general correction and agreement check process has been described in Section \ref{sec:Information_reconciliation}.
Specifically, in this paper, we adopt a BCH code that can correct 20\% mismatch \cite{TVT1}. In the diversity block $k$, Bob selects a code $c$ from the BCH set and calculates $s_{k}=\mathrm{XOR}(K_{B,k},c)$. $s_{k}$ is then transmitted to Alice through the public channel. Alice corrects the disagreed key bits and obtain $K_{A^{'},k}$ using $s_{k}$. After error correction, CRC is employed to confirm the key agreement. If the CRC check values $v_{A,k}$ and $v_{B,k}$ are identical, Alice and Bob agree on the same key, i.e., $K_{A^{'},k}=K_{B,k}$. 
Then Alice will send $\mathrm{IRC}_{k}$ to Bob to indicate whether the keys match. If the keys match, $\mathrm{IRC}_{k}=1$; otherwise, $\mathrm{IRC}_{k}=0$. 

\subsubsection{Guard Band Calibration}
The selected quantization parameters may still result in KDR higher than 20\% due to measurement noises and when their channel variations deviate from the training RSSI dataset significantly. A smaller $\alpha$ generally yields a higher KDR. When the KDR of a diversity block is above 20\%, then the $\alpha$ is likely to be underestimated by the linear regression model. In this case, we propose a guard band parameter calibration approach to decrease the KDR in this diversity block. We increase $\alpha$ by a calibration value $\Delta \alpha$ and perform quantization for this diversity block again using the new guard band parameter $\alpha + \Delta \alpha$. 
This calibration can be repeated until the disagreed key bits of the diversity block are corrected. However, more calibrations lead to higher communication overhead due to more information exchange between Alice and Bob. Let $N_{\mathrm{C}}$ denote the number of calibrations. If the disagreed key bits cannot be corrected within $N_{\mathrm{C}}$ calibrations, the key sequences in this diversity block will be dropped. 

\begin{figure}[!t]
 \centering
 \subfloat[]{\includegraphics[width=3.4in]{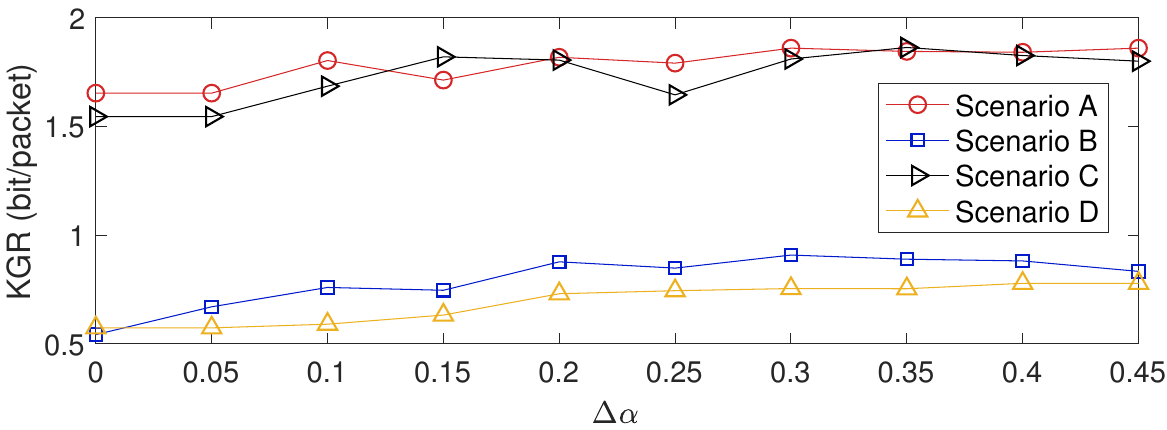}} \\
 \subfloat[]{\includegraphics[width=3.4in]{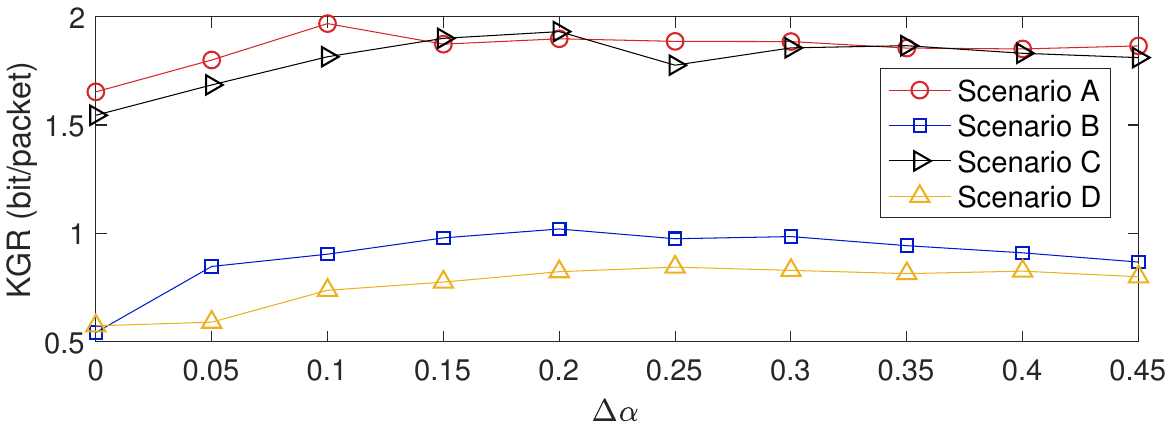}}
\caption{Impact of calibration value. (a) Single calibration. (b) Double calibration.}
\label{calibration}
\end{figure}

Fig. \ref{calibration} shows the impact of guard band parameter calibration on KGR using the RSSI dataset presented in Section~\ref{sec:RSSI_Dataset}. It is observed that KGR first increases and then {slightly decreases} with $\Delta\alpha$ for both single calibration ($N_C = 1$) and double calibration ($N_C = 2$), while double calibration can yield a higher KGR than single calibration. In this paper, we set $\Delta\alpha=0.2$ as it performs well under various scenarios.

 After quantization and information reconciliation for all the diversity blocks, Alice and Bob get key sequences $K_{A^{'}}$ and $K_{B}$, respectively.

\subsection{Privacy Amplification}
 We employ the SHA256 hash function on the key sequences $K_{A^{'}}$ and $K_{B}$ to remove the leaked information and obtain the final secret  keys.

\section{Experimental Evaluation}
\label{PERFORMANCE EVALUATION}

\subsection{Experimental Setup}
To evaluate the performance of our proposed AQ-KG protocol in the online deployment phase, we carried out six new tests in different environments or using different devices, as listed in Table \ref{tab1}. The LoRa parameter configurations are the same as explained in Section~\ref{sec:testbed}.

\begin{figure}[!t]
 \centering
 \subfloat[]{\includegraphics[height=1.6in]{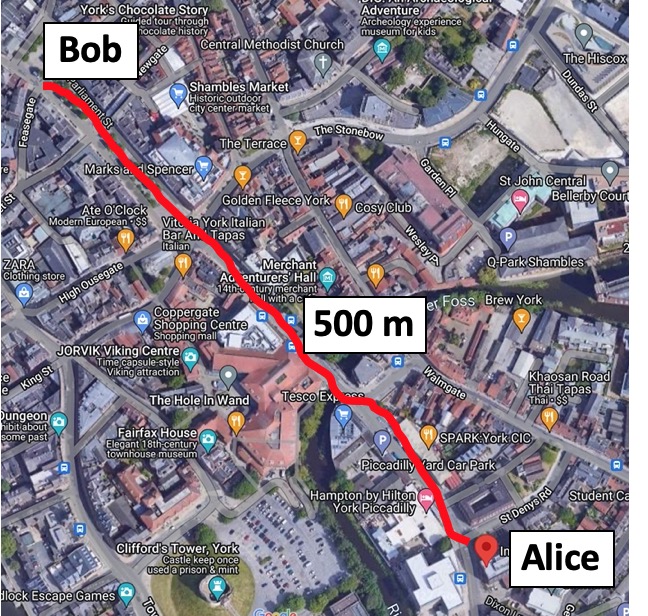}}
 \subfloat[]{\includegraphics[height=1.6in]{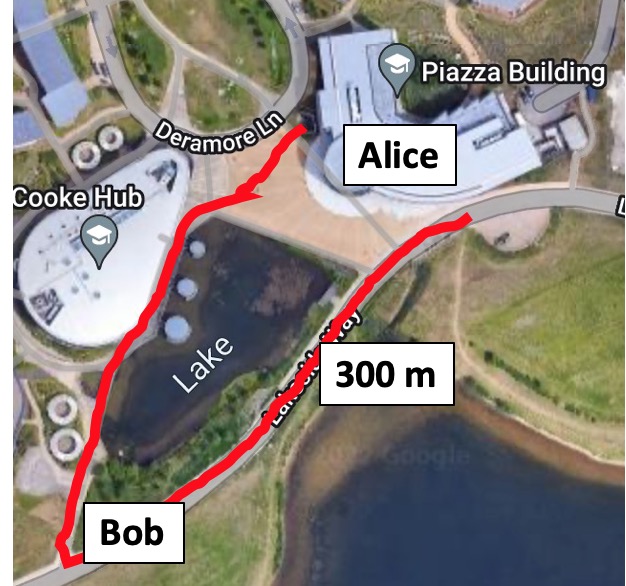}}
\caption{The trajectory of Bob. (a) Test A and C.  (b) Test E.}
 \label{trajectory}
\end{figure}

\begin{figure}[!t]
 \centering
 \subfloat[]{\includegraphics[width=3.4in]{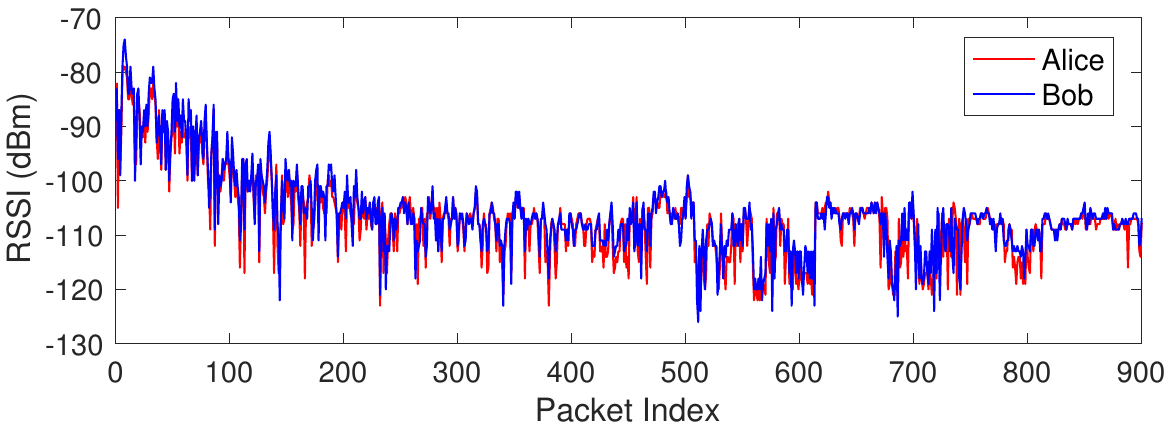}}  \\
 \subfloat[] {\includegraphics[width=3.4in]{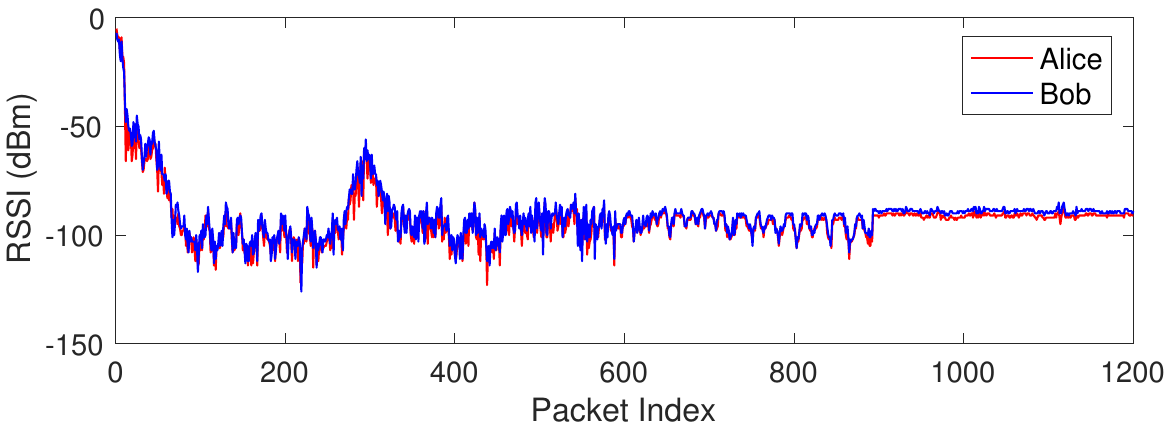}}
\caption{RSSI measurements of Alice and Bob. (a) Test A. (b) Test B.}
\label{Test_data}
\end{figure}

New environments and devices can be used to evaluate the generalization ability of the proposed AQ-KG protocol.
Tests A-D were conducted in the city center of York, U.K, while Test E and F were conducted in the suburban area of York. In all the tests, Alice used Dragino LoRa Shield v1.4. In Test C and D, Bob was replaced with Dragino LoRa/GPS Shield v1.3, as shown in Fig. \ref{device}(b). In Test A and C, Bob moved in a straight line to the city center,  at a walking speed, and then stayed in the city center, as shown in Fig. \ref{trajectory}(a). In Test E, Bob walked around the campus and stayed by the lake for a while, as shown in Fig. \ref{trajectory}(b). In Test B, D, and F, Bob moved in the same building as Alice at a walking speed and then remained stationary. The collected RSSI values of Alice and Bob in Test A and B are exemplified in Fig. \ref{Test_data}, while those collected in other tests are not displayed due to limited space.
\begin{table}[!t]
  \centering
  \caption{Experimental Setup}
  \scalebox{1}{
    \begin{tabular}{|l|L{1.6cm}|L{2.2cm}|L{2.7cm}|}
    \hline
    \textbf{Test} & \textbf{Environment} & \textbf{Alice} & \textbf{Bob} \\
  \hline
    \textbf{A} & Outdoor & Stationary, residential building & Urban, LoRa Shield v1.4 \\
  \hline
    \textbf{B} & Indoor & Stationary, residential building & Residential building, LoRa Shield v1.4 \\
  \hline
    \textbf{C} & Outdoor & Stationary, residential building & Urban, LoRa/GPS Shield v1.3 \\
  \hline
    \textbf{D} & Indoor & Stationary, residential building & Residential building, LoRa/GPS Shield v1.3 \\
 \hline
    \textbf{E} & Outdoor & Stationary, university building & Suburban, LoRa Shield v1.4 \\
  \hline
    \textbf{F} & Indoor & Stationary, university building & University building, LoRa Shield v1.4 \\
  \hline
    \end{tabular}}%
  \label{tab1}%
\end{table}%

\subsection{Comparison with Benchmark Methods}
We compare our proposed AQ-KG protocol with two state-of-the-art quantization algorithms, namely differential quantization \cite{TVT1} and fixed quantization \cite{IOTJ1,IOTJ2,TMC2} schemes.  Differential quantization determines key bits based on the differential value between adjacent RSSI values \cite{TVT1}. To maximize the KGR of differential quantization, we set the RSSI resolution as $0$. In the fixed quantization scheme, the quantization parameters are fixed throughout the quantization process, which is used in most previous works. We iteratively search for the optimal combination of $m$ and $\alpha$ that maximizes the KGR to 
evaluate the performance upper bound of the fixed quantization scheme. Note that the search for optimal parameters is not available in practice as it requires both Alice's and Bob's RSSI data. 

\begin{figure}[!t]
\centerline{\includegraphics[width=3.4in]{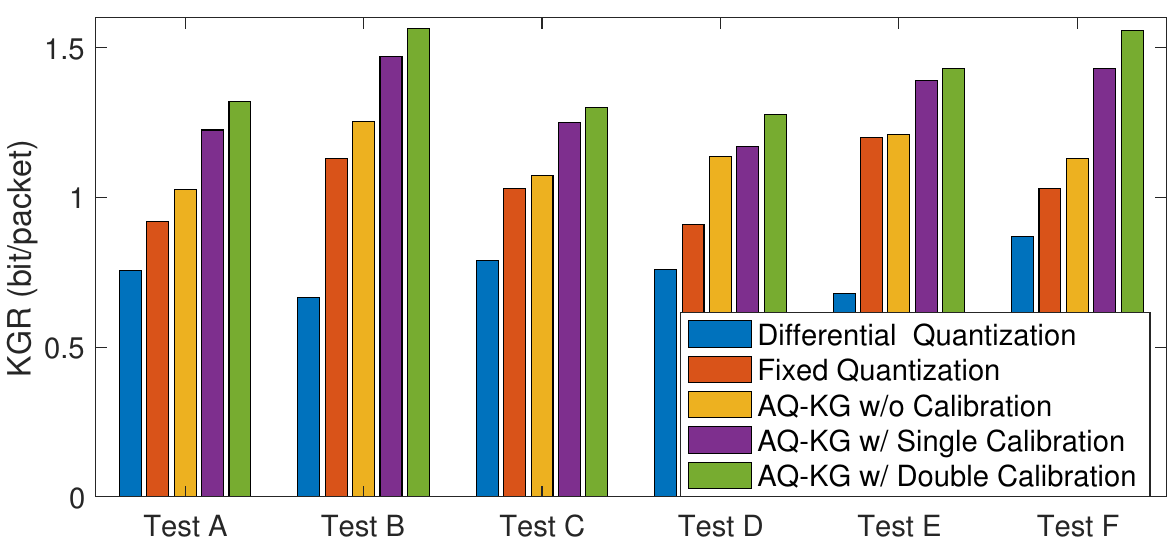}}
\caption{Performance comparison.}
\label{comparison_total}
\end{figure}

 From Fig. \ref{comparison_total}, we can see that the proposed adaptive quantization scheme achieves higher KGR than differential quantization and fixed quantization in all the tests, indicating that the proposed scheme is applicable to different environments and devices.  Moreover, the proposed quantization parameter calibration scheme further improves KGR by adjusting the guard band parameters underestimated by the adaptive quantization scheme. For instance,  KGR is improved by 28.5\% and 24.7\% after double calibration  compared to no calibration in Test A and B, respectively. Among the tests performed, the proposed AQ-KG protocol can achieve up to 2.35$\times$ and 1.51$\times$ KGR gains compared to differential quantization and fixed quantization schemes, respectively.

\begin{figure}[!t]
\centerline{\includegraphics[width=3.4in]{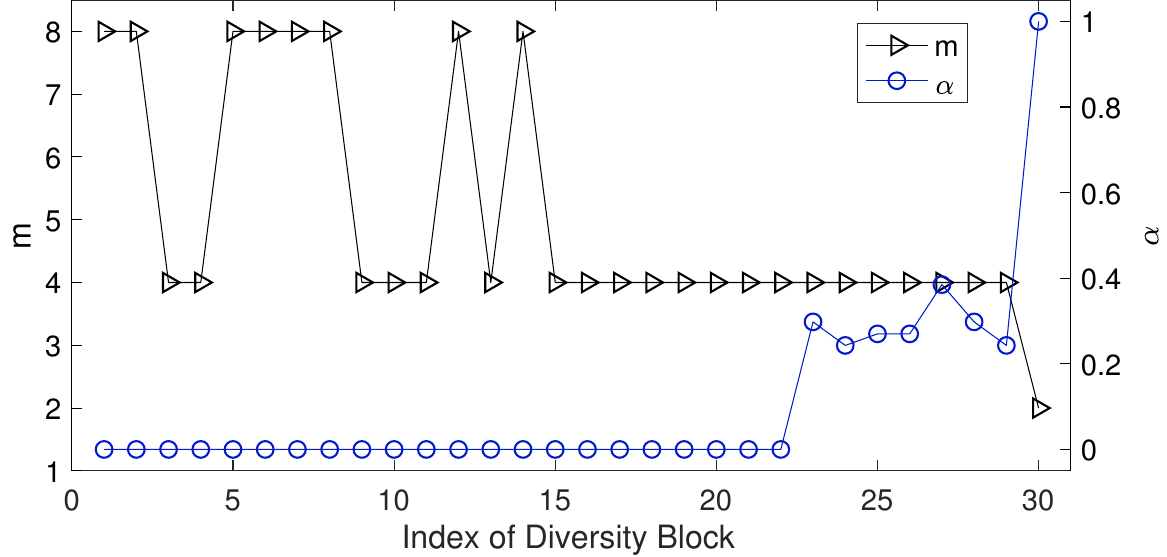}}
\caption{Quantization parameters determined by the proposed adaptive quantization scheme for  Test B.}
\label{parameter_change}
\end{figure}

 Fig. \ref{parameter_change} shows the quantization parameters {for each diversity block} obtained by our proposed adaptive quantization scheme in Test B. It can be seen that our proposed scheme can dynamically adjust the  quantization parameters according to the channel variations. The adaptive quantization parameters of other tests are not presented for brevity.
 
\begin{table}[t]
{
\caption{P-Values of NIST Test}
\begin{center}
\scalebox{0.98}{
\begin{tabular}{|c|c|c|c|c|c|c|}
\hline  
\textbf{Test}&\textbf{\tabincell{c}{A}}&\textbf{\tabincell{c}{B}}&\textbf{\tabincell{c}{C}}&\textbf{\tabincell{c}{D}}&\textbf{\tabincell{c}{E}}&\textbf{\tabincell{c}{F}} \\
\hline  \hline
\tabincell{c}{Frequency} & 0.215 & 0.595 & 0.859 & 0.376 & 0.723 & 0.859\\
\hline 
\tabincell{c}{Block frequency}  & 0.215 & 0.595 & 0.859 & 0.376 & 0.723 & 0.859\\
\hline 
Runs &  0.5 &  0.741 & 0.857 & 0.188 & 0.991 & 0.997\\
\hline 
\tabincell{c}{Longest run of 1s} & 0.682 &0.167 & 0.265 & 0.022 & 0.366 & 0.752\\ 
\hline
DFT & 0.871 & 0.33 & 0.33 & 0.871  & 0.157 & 0.516\\
\hline 
Serial & 0.498 & 0.978 & 0.498 & 0.999 & 0.498 & 0.498\\
\hline 
\tabincell{c}{Appro. entropy} & 0.41 & 0.822 & 0.984 & 0.353 & 0.939 & 0.984\\
\hline 
 \tabincell{c}{Cum. sums (fwd)} & 0.103 & 0.369 & 0.949 & 0.574 & 0.892 & 0.818\\
\hline 
 \tabincell{c}{Cum. sums (rev)} & 0.43 & 0.818 & 0.984 & 0.103 & 0.737 & 0.654\\
\hline 
\end{tabular}}
\label{tab2}
\end{center}}
\end{table} 

\subsection{Randomness Analysis}
We used the NIST randomness test suite to evaluate the randomness of generated keys by our proposed AQ-KG protocol. For each 128-bit key sequence, we observe that the P-values returned are greater than 0.01.  Hence, all the generated 128-bit key sequences pass the NIST tests, i.e., NFR=0. We present the P-values of one of the 128-bit key sequences in each test in Table \ref{tab2}.

\section{Security Analysis}
\label{sec:security_analysis}
A widely considered attack model in
physical layer key generation is passive eavesdropping where the eavesdropper monitors the information exchange between the legitimate nodes, possesses knowledge of the whole key generation protocol and tries to generate the same key as legitimate nodes based on its own channel observations \cite{Access1}. Fortunately, when the eavesdropper is located far away from both legitimate nodes, the channel observed at the eavesdropper is  uncorrelated with the channel between legitimate nodes \cite{TIFS1}. Moreover, it has been experimentally validated in \cite{IOTJ1, IOTJ2} that the passive eavesdropper cannot establish the same key using LoRa devices.

Compared with the quantization methods in the literature, our proposed adaptive multi-level quantization scheme requires the transmission of quantization parameters that would be received by the eavesdropper, but this leaks no information about the channel measurements of legitimate nodes. Similarly, the transmission of IRC during a guard band parameter calibration provides no information about the generated keys. Hence, the proposed AQ-KG protocol can establish secure keys as long as the eavesdropping channels are uncorrelated with the legitimate channels.

Physical layer key generation can also be vulnerable to man-in-the-middle (MiM) attacks, in which the active attacker intercepts the information transmitted between the legitimate nodes and inject new information. A countermeasure is to authenticate the user’s identity and deny access to malicious users \cite{WC1}. Recently,  radio frequency fingerprint identification has emerged as an effective technique  to authenticate LoRa devices \cite{robyns2017physical, shen9448147,al2021deeplora,shen2022towards}. {An alternative solution is to introduce a reconfigurable antenna, which can change the channel state after each channel probing, thus preventing the
attacker from predicting the channel state of packet injection
attacks \cite{pan2021man}.}


\section{Conclusions}
\label{Conclusions}
In this paper, we have investigated physical layer key generation for LPWANs using LoRa devices. In particular, we proposed an adaptive  multi-level quantization scheme to fully exploit channel randomness by dynamically selecting quantization parameters according to channel variations. Based on the adaptive quantization scheme, we proposed an AQ-KG protocol that includes a guard band parameter calibration scheme to adjust guard band parameters when information reconciliation fails.  We conducted extensive real-world channel measurements in various scenarios. Our experimental results 
demonstrated that the proposed AQ-KG protocol can achieve up to 2.35$\times$ and 1.51$\times$ KGR gains compared to the state-of-the-art differential quantization and fixed quantization schemes, respectively. {The extension of the proposed adaptive quantization scheme to other communication techniques such as WiFi and Bluetooth will be of interest in our future work.}



\end{document}